\documentclass[journal]{IEEEtran}
\IEEEoverridecommandlockouts
\usepackage{cite}
\usepackage{amsmath,amssymb,amsfonts}
\usepackage{algpseudocode}
\usepackage{graphicx}
\usepackage{textcomp}
\usepackage{subfigure}
\usepackage{xcolor}
\usepackage{optidef}
\usepackage{algorithm}
\usepackage{epstopdf}

\DeclareMathOperator{\diag}{diag}
\let\oldexist\exists
\renewcommand{\exists}{\oldexist \: }

\let\oldforall\forall
\renewcommand{\forall}{\oldforall \, }
\def\BibTeX{{\rm B\kern-.05em{\sc i\kern-.025em b}\kern-.08em
    T\kern-.1667em\lower.7ex\hbox{E}\kern-.125emX}}
\begin{document}

\title{User Equipment Assisted Localization for 6G Integrated Sensing and Communication
\thanks{X. Guo, Q. Shi, S. Zhang, and L. Liu are with the Department of Electrical and Electronic Engineering, The Hong Kong Polytechnic University, Hong Kong SAR, China (e-mails: \{xianzhen.guo,qin-eie.shi\}@connect.polyu.hk, \{shuowen.zhang,liang-eie.liu\}@polyu.edu.hk).}
\thanks{C. Xing is with the School of Information and Electronics, Beijing Institute of Technology, Beijing 100081, China (e-mail: xingchengwen@gmail.com).} 
\thanks{The materials in this paper have been presented in part at the {\it IEEE Int. Conf. Acoust., Speech Signal Process. (ICASSP)}, April 2024 \cite{Guo2024}.}}

\author{\IEEEauthorblockN{Xianzhen Guo, Qin Shi, Shuowen Zhang, Chengwen Xing, and Liang Liu}}

\maketitle

\begin{abstract}
This paper investigates user equipment (UE) assisted device-free networked sensing in the sixth-generation (6G) integrated sensing and communication (ISAC) system, where one base station (BS) and multiple UEs, such as unmanned aerial vehicles (UAVs), serve as anchors to cooperatively localize multiple passive targets based on the range information.
Three challenges arise from the above scheme. First, the UEs are not perfectly synchronized with the BSs. Second, the UE (anchor) positions are usually estimated by the Global Positioning System (GPS) and subject to unknown errors. Third, data association is challenging, since it is hard for each anchor to associate each rang estimation to the right target under device-free sensing. We first tackle the above three challenges under a passive UE based sensing mode, where UEs only passively hear the signals over the BS-target-UE paths. 	
A two-phase UE assisted localization protocol is proposed. In Phase I, we design an efficient method to accurately estimate the ranges from the BS to the targets and those from the BS to the targets to the UEs in the presence of synchronization errors between the BS and the UEs. In Phase II, an efficient algorithm is proposed to localize the targets via jointly removing the UEs with quite inaccurate position information from the anchor set and matching the estimated ranges at the BS and the remaining UEs with the targets. Next, we also consider an active UE based sensing mode, where the UEs can actively emit signals to obtain additional range information from them to the targets. We show that this additional range information can be utilized to significantly reduce the complexity of Phase II in the aforementioned two-phase localization protocol. Numerical results show that our proposed UE assisted networked sensing scheme can achieve very high localization accuracy. 
\end{abstract}


\section{Introduction}
\subsection{Motivation}\label{subsec:motivation}
Thanks to the sufficient bandwidth at the millimeter wave (mmWave) band and the terahertz (THz) band as well as the large aperture array brought by the massive multiple-input multiple-output (MIMO) technique, the sixth-generation (6G) cellular network will be able to provide high-resolution sensing services \cite{liu2022integrated,zheng2019,zhang2021,Hassanien,liuan2022}. Recently, the International Telecommunication Union Radiocommunication Sector (ITU-R) Study Group 5 has identified integrated sensing and communication (ISAC) as one of the six usage scenarios for the 6G cellular network in ``The ITU-R Framework for IMT-2030''. One notable advantage of 6G-enabled sensing over radar sensing lies in the potential for large-scale networked sensing \cite{liu2024maga, Yang23, win09, shi2022}. Specifically, the widely deployed base stations (BSs) in the cellular network are inter-connected by the fronthaul/backhaul networks and can share their local information for better sensing performance. This philosophy is similar to that behind the cooperative communication techniques such as networked MIMO, cloud radio access network, etc.

In addition to the BSs, the Third Generation Partnership Project (3GPP) recently recommended to adopt the user equipments (UEs) as anchors to perform certain sensing tasks \cite{3GPP}. Compared to the BSs, the deployment density for the UEs is much higher, making it easier to achieve ubiquitous sensing. For example, it has been popularly suggested that the unmanned arial vehicles (UAVs) should be equipped with cellular signal transceivers to become the aerial UEs subscribed to the system, such that they can exchange crucial control information with the BSs reliably throughout their flight period \cite{Zhang19UAV}. For the future 6G-connected UAVs, they can thus fly to particular hotspots of interests, extract sensing information about their surrounding environment from their received echo signals, and feed back these sensing information to a BS via wireless channels, which can perform networked sensing based on all collected information.

Motivated by the above, this paper considers a novel UE assisted networked sensing architecture, where multiple UEs help a BS to localize multiple passive targets. Under this framework, the BS and all the UEs share their sensing information for better localization performance. However, different from the conventional networked sensing architecture where all the anchors are the BSs \cite{shi2022}, the UE assisted networked sensing architecture gives rise to new challenges. 
\begin{itemize}
\item Challenge 1: Synchronization. In practice, the UEs and the BS are not perfectly synchronized. The sampling timing offsets (STOs) between the BS and the UEs will affect the accuracy for estimating the propagation delay from the transmitting anchor to a different receiving anchor via a target. 
\item Challenge 2: UE Position Uncertainty. Different from the static BS, the UEs are mobile and their positions have to be estimated by the Global Positioning System (GPS). However, in practice, the GPS may provide very erroneous position information to some UEs, e.g., when they are surrounded by high buildings. Therefore, we should be able to dynamically identify these UEs and remove them from the anchor set to maintain the localization performance under networked sensing. 
\item Challenge 3: Data Association. Under device-free networked sensing, how to match each echo signal received by an anchor to the right target that generates this signal, i.e., data association, is a long-standing challenge \cite{Mahler07,liu2024maga,shi2022}. 
\end{itemize}In this paper, we aim to tackle the above challenges to enable UE assisted sensing.
\vspace{-2mm}

\subsection{Prior Work}
UE assisted sensing has been investigated in the literature about mapping or simultaneous localization and mapping (SLAM) \cite{Guidi2016, BaqueroBarneto2022, Gentner2016, Yang2022}. Specifically, in \cite{Guidi2016} and \cite{BaqueroBarneto2022}, the UEs moving in predetermined trajectories are utilized to construct a map of its surrounding environment based on the signals transmitted by themselves, while \cite{Gentner2016} and \cite{Yang2022} aim to perform SLAM using UEs whose position information is unknown and needs to be joint estimated with the states of objects in the environment. Different from the existing works which utilize dedicated UEs for sensing \cite{Guidi2016, Gentner2016} or are limited in the  indoor environment \cite{BaqueroBarneto2022, Yang2022}, this paper aims to explore the possibility of using the existing UEs in the cellular network, which are primarily designed for communication, to perform sensing. In particular, we are interested in devising a novel UE assisted sensing architecture based on the communication signals in the cellular network, where the UEs and the BS can work together to provide ubiquitous sensing services.
 
In the literature, some works have been done to tackle Challenges 1, 2, and 3 listed in the above under different scenarios. Specifically, as to the synchronization issue between the UE and the BS, \cite{Li2020} assumes that the UEs (receivers) have the same STO with respect to the BS (transmitter) and eliminates the STO utilizing the delay difference at different UEs.
However, in practice, because the UEs are not perfectly synchronized, they have different STOs to the BS. Furthermore, in \cite{Ni2021}, the STO between the transmitter and the receiver is mitigated by correlating the signals received at different antennas at the receiver. However, in practice, many UEs are only equipped with one antenna, and the above algorithm thus cannot be applied.

Besides synchronization, there are a large number of prior works in wireless sensing networks (WSN) investigating how to mitigate the effect of anchor position uncertainty on localization \cite{shiqingjiang2010, Ghari2019, lui2009, ghasem2014, angjeli2015}. In these works, the exact covariance matrices of the position errors are assumed to be known. Then, these information is used in defining the maximum likelihood (ML) problem to localize the targets, such that the residuals associated with the anchors with smaller position errors are assigned with larger weights. As a result, the anchors with little position uncertainties can play a more significant role in localizing the targets. However, in practice, it is difficult to obtain the covariance matrices of the errors in UE positions estimated by GPS. Thus, these methods cannot be used in our paper, in which the position error distribution is not available. 

At last, the data association issue \cite{Mahler07}, which is challenging in multi-source multi-target systems, including our considered UE-assisted sensing system, has been widely studied in the literature. Many efficient algorithms, such as the nearest neighbor (NN) algorithm\cite{BarShalom2009}, the probabilistic data association filter (PDAF) \cite{PDA2004}, and the joint probabilistic data association filter (JPDAF) \cite{JPDA1983} have been proposed. Particularly, when all the anchors are the BSs, \cite{Shi2024} and \cite{Molisch2014} proposed joint data association and localization algorithms to estimate the positions of multiple targets. However, under our considered UE assisted sensing network, data association is coupled with Challenge 1, i.e., synchronization, and Challenge 2, i.e., anchor position uncertainty. A new algorithm that can tackle all these challenges together is needed.  
\begin{figure}
    \centering
    \includegraphics[width=0.9\linewidth]{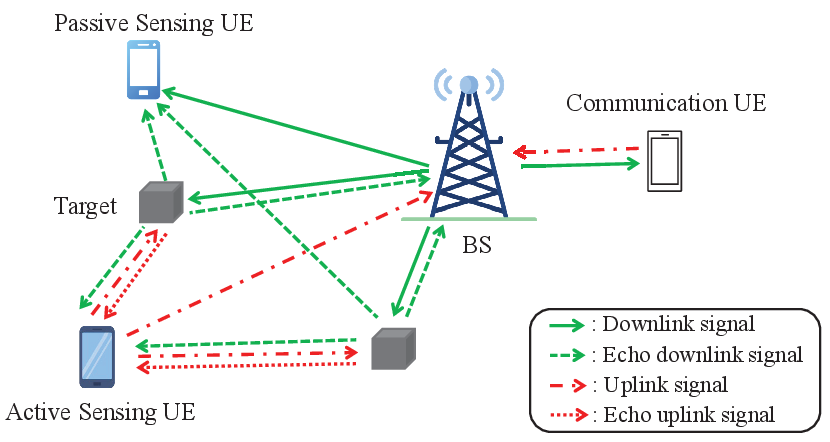}
    \caption{UE assisted ISAC network: Some UEs just communicate with the BS in the uplink/downlink. The other UEs can aid the BS to perform sensing. Specifically, some idle UEs can passively hear the echo signals reflected by the targets to perform passive sensing. On the other hand, some UEs can utilize their uplink signals to actively sense the targets. Because uplink/downlink communication techniques are quite mature, this paper mainly considers how the UEs can assist the BS to localize the targets.}
    \label{sysmodel}
    \vspace{-5mm}
\end{figure}
\subsection{Main Contribution} 
As shown in Fig.~\ref{sysmodel}, this paper considers a networked sensing system where one BS and multiple UEs cooperatively localize multiple passive targets. \footnote{In 6G ISAC networks, some other UEs may just communicate with the BS in the uplink/downlink, as shown in Fig.~\ref{sysmodel}. However, communication techniques are quite mature now. Therefore, in this paper, we do not consider the communication part in the ISAC networks.} Two sensing modes will be studied. Specifically, under the passive UE based sensing mode, the BS transmits the orthogonal frequency division multiplexing (OFDM) communication signals in the downlink, while the UEs only passively hear the echo signals scattered by the targets to measure the BS-target-UE ranges. Such a type of range information is then fed back by the UEs to the BS, which can localize the targets based on the global range information. Under the active UE based sensing mode, besides the BS, the UEs also actively emit uplink signals at a different frequency band to probe the environment. Under this mode, each UE can estimate the BS-target-UE range and the target-UE ranges based on the echo signals received at the downlink band and the uplink band, respectively. Then, these two types of range information are fed back to the BS for performing localization. The contributions of this paper are summarized as follows.

First, we devise a novel method to estimate the STOs between the BS and the UEs such that the propagation delay between the imperfectly synchronized BS and UE via some target can be accurately estimated. Specifically, we first estimate the effective propagation delay from the BS to each UE over the LOS path, which is the superposition of the true propagation delay and the STO between the BS and each UE, based on the OFDM channel estimation technique. Then, we calculate the true propagation delay from the BS to each UE based on their locations. Last, the difference between the effective propagation delay and the true propagation delay of any LOS path is the STO between the BS and the corresponding UE. After estimating the STOs, the propagation delay from the BS to each target to each UE is correctly estimated. 

Second, we devise an efficient scheme to localize the targets under the passive UE based localization strategy. We start with the special case when only one target exists (data association is not required in this case because all the range information belongs to this target) as in \cite{Guo2024}. Under this case, we are able to identify the ineffective UEs whose positions estimated by the GPS are quite erroneous based on the outlier detection technique \cite{li2006, Guvenc2009}. Specifically, we propose an iterative algorithm, while in each iteration, one ineffective UE is identified and removed from the UE set if the removal of it can lead to the maximum reduction in the residue of localizing the target. This process continues until no significant residue reduction is found by removing any UE from the anchor set. Next, we consider the general case with multiple targets, where data association (not considered in \cite{Guo2024}) and UE position uncertainty are closely coupled. A key observation is that an effective/ineffective UE is beneficial/harmful to localize all the targets. Therefore, we propose to decouple ineffective UE removal and data association as follows. First, we design an efficient algorithm that can select a good target such that the effective UE set can be obtained based on our proposed method under the single-target case. Second, we adopt this effective UE set when localizing the other targets, while we only need to deal with the data association issue based on the method proposed in \cite{shi2022}.

Last, under the active UE based sensing mode, which is not considered in \cite{Guo2024}, we show that the additional information about the target-UE ranges, obtained from uplink signals, enables a more efficient localization method. 
A key observation is that the range of the path from the BS via a target to a UE should equal the sum of the distance from the BS to the target and the distance from the target to the UE. Leveraging this observation, we can eliminate some ineffective UEs and some infeasible data association solutions for each target. Based on this, we modify certain steps in the proposed method for passive UE based sensing mode, resulting in improved localization performance and reduced computational complexity.

\subsection{Organization}
The rest of the paper is organized as follows. Section \ref{sec:system model} introduces the system model of UE assisted networked sensing. Sections \ref{sec:Localization Method under Passive UE based Sensing Mode} and \ref{sec:Localization Method under Active UE based Sensing Mode} propose efficient methods to enable UE assisted networked sensing under the passive UE based and the active UE based sensing modes, respectively. Numerical results are provided in Section \ref{sec:Numerical Results}. Finally, Section \ref{sec:Conclusions} concludes this paper.

\section{System Model}
\label{sec:system model}
We consider a 6G network based sensing system consisting of one BS, $M$ UEs, denoted by $\mathcal{M} = \{1,\dots, M\}$, and $K$ targets, denoted by $\mathcal{K} = \{1,\dots,K\}$, as illustrated in Fig.~\ref{sysmodel}. The system is working in the frequency-division duplexing (FDD) mode such that the BS and the UEs can potentially transmit radio signals at the same time but over different frequency bands. 
Let $\boldsymbol{b} \in \mathbb{R}^{2 \times 1}$, $\boldsymbol{u}_m \in \mathbb{R}^{2 \times 1}$, and $\boldsymbol{a}_k \in \mathbb{R}^{2 \times 1}$ denote the 2D coordinates of the BS, the $m$-th UE, $\forall m \in \mathcal{M}$, and the $k$-th target, $\forall k \in \mathcal{K}$, respectively. 
Then, we can define $d^{\rm BT}_k = \|\boldsymbol{b}-\boldsymbol{a}_k\|$ and $d^{\rm UT}_{m,k} = \|\boldsymbol{u}_m-\boldsymbol{a}_k\|$ as the true distance between the BS and target $k$ and that between UE $m$ and target $k$, respectively, $\forall m, k$. 
Moreover, the sum of the distance between the BS and target $k$ and that between UE $m$ and target $k$ is
\begin{equation}\label{sumdistance}
    d^{\rm BTU}_{m,k} = d^{\rm BT}_k + d^{\rm UT}_{m,k}, \forall m, k.
\end{equation} 

Under the above system, the BS and the UEs cooperatively serve as the anchors to localize the targets. 
In practice, the BS's location is fixed and perfectly known,  while the UEs' locations are typically estimated via GPS and subject to estimation errors. Define $\hat{\boldsymbol{u}}_m = \boldsymbol{u}_m + \Delta \boldsymbol{u}_m$ as the estimated position of UE $m$ made by GPS, where $\Delta \boldsymbol{u}_m$ denotes the unknown position estimation error, $\forall m$. Because the location of UE $m$ is known as $\hat{\boldsymbol{u}}_m$, instead of $\boldsymbol{u}_m$, any estimations of $d^{\rm UT}_{m,k}$ and $d^{\rm BTU}_{m,k}$ are believed to be the values of $\|\hat{\boldsymbol{u}}_m-\boldsymbol{a}_k\|$ and $\|\boldsymbol{b}-\boldsymbol{a}_k\|+\|\hat{\boldsymbol{u}}_m-\boldsymbol{a}_k\|$, respectively, $\forall m, k$. 

In this paper, depending on the role of the UEs, we consider two sensing modes - passive UE based sensing mode and active UE based sensing mode. Under the first mode, only the BS emits the OFDM signals for the sensing purpose, while the UEs just passively listen to the echo signals reflected by the targets to perform localization. Under the second mode, besides the BS, the UEs also emit the radio signals to actively probe the environment. Since the OFDM cellular communication technology is quite mature, in the rest of this paper, we mainly study how to leverage the OFDM communication signals for sensing the targets with the BS and the UEs serving as anchors in our considered UE-assisted sensing system.

\subsection{Passive UE based Sensing Mode}\label{sec:Passive UE based Sensing Mode}
Under the passive UE based sensing mode, merely the BS emits the downlink OFDM signals to sense the environment. Let $\mathcal{N}_d = \{1,\dots,N_d\}$ denote the set of sub-carriers in the downlink. Moreover, let $\boldsymbol{s} = [s_{1},\dots,s_{N_d}]^T$ denote the frequency-domain OFDM symbol sent from the BS, where $s_{n}$ denotes the unit-power transmitted sample at the $n$-th sub-carrier in the downlink, $\forall n \in \mathcal{N}_d$. Then, the corresponding time-domain OFDM signal of the BS is $\boldsymbol{\chi} = \left[\chi_{1},\dots,\chi_{N_d}\right]^T = \boldsymbol{W}_d^H\sqrt{p_0}\boldsymbol{s},$ where $\chi_{n}$ denotes the $n$-th sample, $p_0$ denotes the transmission power at the BS, and $\boldsymbol{W}_d \in \mathbb{C}^{N_d\times N_d}$ denotes the discrete Fourier transform (DFT) matrix.
    Note that the duration of each downlink OFDM sample period is $1/ N_d\Delta f_d$ seconds (s), where $\Delta f_d$ in Hz denotes the downlink OFDM sub-carrier spacing. After inserting the cyclic prefix (CP) consisting of $Q_d$ OFDM samples, the time-domain signal transmitted by the BS over one downlink OFDM symbol period is given by $\bar{{\boldsymbol{\chi}}} = [\bar{\chi}_{-Q_d},\dots,\bar{\chi}_{-1},\bar{\chi}_{0},\dots,\bar{\chi}_{N_d-1}]$, where $\bar{\chi}_{n} = \chi_{N_d+n+1}$ denotes the CP when $n \leq 0$, and $\bar{\chi}_{n} = \chi_{n+1}$ denotes the useful signal when $n > 0$.

Define $\boldsymbol{h}^{\rm BTB} = [h_0^{\rm BTB},\dots,h_{L-1}^{\rm BTB}]$ as the $L$-tap baseband equivalent channel from the BS via the targets to the BS, where $L$ denotes the maximum number of detectable paths.  
Note that $h^{\rm BTB}_l \neq 0$ holds if and only if there exists a target such that the delay of the propagation path from the BS via this target to the BS is of $l$ OFDM sample periods. 
Then, the received signal at the BS in the $n$-th OFDM sample period, which arises from the echoes of the targets, is expressed as 
\begin{equation}\label{bs2bs}
y_{n} = \sum_{l=0}^{L-1}h_l^{\rm BTB}\bar{\chi}_{n-l} + z_{n}, \forall n,
\end{equation}
where $z_{n}\sim \mathcal{CN}(0,\sigma^2)$ denotes the noise at the BS in the $n$-th downlink OFDM sample period, with $\sigma^2$ denoting the average noise power.  

The UEs can passively listen to their received echo signals and help the BS perform localization. Specifically, each UE can receive the downlink signals emitted by the BS via two types of links: the \textit{direct link} from the BS to the UE and the \textit{cascaded links} from the BS to the targets to the UE. 
In practice, the UEs are not perfectly synchronized in time with the BS. Define the sampling timing offset (STO) between the BS and UE $m$ as $\tau_m$ in terms of OFDM sample periods, $\forall m$, i.e., if the local clock time at UE $m$ is $t_m$, then that at the BS is $t = t_m + \tau_m$. Specifically, we have $\tau_m > 0$ if the clock time at UE $m$ is earlier, while $\tau_m < 0$ otherwise. 
Let $\boldsymbol{h}_m^{\rm BU} = [h_{m,0}^{\rm BU},\dots,h_{m,L-1}^{\rm BU}]$ denote the $L$-tap multipath channel from the BS via the targets to UE $m$. Note that $h_{m,l}^{\rm BU} \neq 0$ holds if and only if $l$ is the delay (in terms of OFDM sample periods) of the propagation either from the BS to UE $m$ or from the BS to some target to UE $m$. 
Then, the received time-domain downlink OFDM signal at UE $m$ in the $n$-th OFDM sample period is given by \footnote{In practice, both STO and carrier frequency offset (CFO) may affect the OFDM systems. Because there are plenty of advanced CFO estimation techniques proposed for OFDM communication \cite{Morelli2007}, in this paper, we mainly focus on how to mitigate the effect of STO on 6G ISAC systems.}
\begin{equation}\label{bs2ue}
    y_{m,n}^{\rm DL} = \sum_{l=0}^{L-1} h_{m,l}^{\rm BU} \bar{\chi}_{n-l-\tau_m} + \check{z}_{m,n}, \forall m, n \in \mathcal{N}_d, 
\end{equation}
where $\check{z}_{m,n}\sim \mathcal{CN}(0,\check{\sigma}_{m}^2)$ denotes the noise at UE $m$ in the $n$-th OFDM sample period, with $\check{\sigma}_{m}^2$ denoting the average noise power.

Under the passive UE based sensing mode, at the downlink frequency band, the system aims to utilize the BS's received signal \eqref{bs2bs} for estimating the BS-target distance and UEs' received signals \eqref{bs2ue} for estimating the BS-target-UE distance to localize the targets. 

\subsection{Active UE based Sensing Mode}\label{sec:Active UE based Sensing Mode}
Under the active UE based sensing mode, besides the BS, the UEs also transmit uplink signals at a frequency band that is different to the downlink frequency band for the sensing purpose. Note that in the fifth-generation (5G) network, besides single-carrier frequency division multiple access (SC-FDMA) signals, it is suggested that OFDM signals can also be transmitted in the uplink \cite{UPOFDM}. In this paper, we assume that the active UEs transmit OFDM signals in the uplink. Specifically, denote the number of the OFDM sub-carriers and the sub-carrier spacing in the uplink as $N_u$ and $\Delta f_u$ Hz, respectively. Moreover, define $\mathcal{N}_u = \{N_d+1,\dots,N_d+N_u\}$ as the set of sub-carriers in the uplink. Note that $\mathcal{N}_d \cap \mathcal{N}_u = \emptyset$, because in the FDD mode, the BS and the UEs transmit at different frequency bands. To reduce the inter-UE interference, different UEs are allocated to orthogonal sub-carriers in the uplink.   
 Let $\mathcal{N}_u^{(m)}$ denote the set of sub-carriers allocated to UE $m$ in the uplink, where $\cup_{m=1}^M \mathcal{N}_u^{(m)} = \mathcal{N}_u$ and $\mathcal{N}_u^{(m)} \cap \mathcal{N}_u^{(\bar{m})} = \emptyset, \forall m \neq \bar{m}$. 
 Let $\boldsymbol{r}_m = [r_{m,1},\dots,r_{m,N_u}]^T$ denote the frequency-domain uplink  OFDM symbol sent from UE $m$, where $r_{m,n}$ denotes unit-power transmitted sample at the $n$-th sub-carrier. Note that $r_{m,n} = 0$ if $n \notin \mathcal{N}_u^{(m)}$. 
 Next, the time-domain OFDM signal generated by UE $m$ is expressed as
\begin{equation}
    \boldsymbol{\gamma}_m = \left [\gamma_{m,1},\dots,\gamma_{m,N_u}\right ]^T = \boldsymbol{W}_u^H \sqrt{p_m} \boldsymbol{r}_m, \forall m, 
 \end{equation}
 where $\gamma_{m,n}$ denotes the $n$-th sample generated by UE $m$, $p_m$ denotes the transmit power at UE $m$, and $\boldsymbol{W}^H_u \in \mathbb{C}^{N_u \times N_u}$ denotes the DFT matrix. 
 After inserting the CP consisting of $Q_u$ OFDM samples, the time-domain signal transmitted by UE $m$ over one OFDM symbol period is given by 
 $\bar{\gamma}_m = \left [\bar{\gamma}_{m,-Q_u},\dots,\bar{\gamma}_{m,-1},\bar{\gamma}_{m,1},\dots,\bar{\gamma}_{m,N_u-1}\right]$, where if $n\leq 0$, $\bar{\gamma}_{m,n} = \gamma_{m,N_u+n+1}$ denotes the CP, and if $n>0, \bar{\gamma}_{m,n} = \gamma_{m,n+1}$ denotes the useful signal. Let $\boldsymbol{h}^{\rm UU}_{u,m} =  [h^{\rm UU}_{u,m,1},\dots,h^{\rm UU}_{u,m,L} ]$ denote the channel for the link from UE $u$ via/not via the targets to UE $m$, $\forall u, m$. Note that $h^{\rm UU}_{u,m,l} \neq 0$ holds if and only if the propagation delay of the path from UE $u$ via/not via the target to UE $m$ is of $l$ uplink OFDM sample periods. 
 In practice, the UEs are not perfectly synchronized with each other. Define the STO between UE $u$ and UE $m$ as $\epsilon_{u,m}=\tau_u-\tau_m$ in terms of OFDM samples. Then, at the uplink frequency band the signal received by UE $m$ in the $n$-th uplink OFDM sample period is expressed as 
\begin{equation}\label{u_signal}
    y_{m,n}^{\rm UL} = \sum_{u=1}^M \sum_{l=0}^{L-1} h^{\rm UU}_{u,m,l} \bar{\gamma}_{u,n-l-\epsilon_{u,m}} + \hat{z}_{m,n}, \forall m,n,
\end{equation}
where $\hat{z}_{m,n}\sim \mathcal{CN}(0,\hat{\sigma}_{m}^2)$ denotes the noise at UE $m$ in the $n$-th OFDM sample period, with $\hat{\sigma}_{m}^2$ denoting the average noise power.

To summarize, under the active UE based sensing mode, each UE can receive echo signals at two different frequency bands. One type of received signals is at the downlink frequency band as shown in \eqref{bs2ue}, which is contributed by the signals emitted by the BS. Another type of received signals is at the uplink frequency band as shown in \eqref{u_signal}, which is contributed by the signals emitted by the UEs. Therefore, the system can utilize the signals received by the BS given in \eqref{bs2bs}, the signals received by the UEs at the downlink frequency band given in \eqref{bs2ue}, and the signals received by the UEs at the uplink frequency band given in \eqref{u_signal} (which are useful to estimate the target-UE distance) to localize the targets.\footnote{The BS can also receive echo signals at the uplink frequency band. However, these signals provide the same information as that provided by the signals received by the UEs at the downlink frequency band given in \eqref{bs2ue}, i.e., the sum of the BS-target distance and the UE-target distance. Therefore, in this paper, we do not utilize the signals received by the BS at the uplink frequency band.} Therefore, the key difference of the active UE based scheme to the passive UE based schemes lies in the availability of the signals given in (\ref{u_signal}). 

In the following, we show how to perform localization under the passive UE based sensing mode and the active UE based sensing mode, respectively.

\section{Localization Method under Passive UE based Sensing Mode}\label{sec:Localization Method under Passive UE based Sensing Mode}

Under the passive UE based sensing mode, we adopt a two-phase sensing protocol for target localization. Specifically, in Phase \uppercase\expandafter{\romannumeral1}, the BS and the UEs first estimate the time-domain channels based on their downlink received signals \eqref{bs2bs} and \eqref{bs2ue}, and then estimate $d_k^{\rm BT}$'s and $d_{m,k}^{\rm BTU}$'s based on the non-zero entries of the estimated channels, respectively. Thus, after Phase \uppercase\expandafter{\romannumeral1}, the BS can obtain a set of range information of the targets, denoted by $\mathcal{D}^{\rm BT} = \{\bar{d}_1^{\rm BT},\dots,\bar{d}_K^{\rm BT}\}$, where $\bar{d}_k^{\rm BT}$ denotes the estimation of $d_k^{\rm BT}$, $\forall k$. 
Moreover, UE $m$ can also obtain a set of range information of the targets, denoted by $\mathcal{D}^{\rm BTU}_m = \{\bar{d}_{m,1}^{\rm BTU},\dots,\bar{d}_{m,K}^{\rm BTU}\}$, where $\bar{d}_m^{\rm BTU}$ denotes the estimation of $d_{m,k}^{\rm BTU}$. 
The main difficulty in Phase \uppercase\expandafter{\romannumeral1} that we need to tackle lies in the estimation of $d_{m,k}^{\rm BTU}$'s since the BS and the UEs are not perfectly synchronized and the estimated propagation delays from the BS to the target to the UEs will be shifted by the unknown STOs $\tau_m$'s.

Then, in Phase \uppercase\expandafter{\romannumeral2}, given the range information obtained in Phase \uppercase\expandafter{\romannumeral1}, 
we will first estimate the range between target $k$ and UE $m$, i.e., ${d}^{\rm UT}_{m,k}$, as
 \begin{equation}\label{hatdut}
   \hat{d}^{\rm UT}_{m,k} = \bar{d}^{\rm BTU}_{m,k} - \bar{d}^{\rm BT}_k, \forall m, k.
 \end{equation}
Then, we will estimate the targets' locations by using the multilateration method based on the following relationship to the anchors
\begin{align}
	\|\boldsymbol{b}-\boldsymbol{a}_k\| &= \bar{d}_k^{\rm BT} + \eta_k, \forall k, \label{sigma}\\
	 \|\hat{\boldsymbol{u}}_m-\boldsymbol{a}_k\| &= \hat{d}^{\rm UT}_{m,k} + \xi_{m,k}, \forall m, k,\label{epsilon}
\end{align}
where $\eta_k$ and $\xi_{m,k}$ denote the corresponding estimation errors.
Note that in $\eqref{epsilon}$, the location of UE $m$ is deemed as $\hat{u}_m$ because only the estimated location is available. 
 However, there are two main challenges under Phase II. The first challenge is the so-called data association issue \cite{shi2022}, which is common in networked device-free sensing \cite{Torrieri1984}. In particular, to localize target $k$, we need to respectively find which elements in  $\mathcal{D}^{\rm BT}$ and $\mathcal{D}_m^{\rm BTU}$ are $\bar{d}^{\rm BT}_k$ and $\bar{d}^{\rm BTU}_{m,k}$, respectively, such that we can estimate its distance to each UE based on \eqref{hatdut} and localize it by applying the multilateration method based on \eqref{sigma} and \eqref{epsilon}. 
The second challenge lies in the imperfect knowledge about the locations of anchors. Specifically, 
the locations of the UEs (anchors) are estimated by GPS and subject to unknown (and maybe significant) errors $\Delta \boldsymbol{u}_m$'s. In other words, if we simply use all the UEs as anchors, it is quite possible that $\xi_{m,k}$'s are very large for some $m$'s and the corresponding estimated location of the target is very inaccurate. 
To summarize, in Phase \uppercase\expandafter{\romannumeral2}, our main job is to select the UEs with accurate estimated positions as anchors and find the corresponding data association solution to perform the multilateration method with high localization accuracy. 
In the following, we provide detailed information about how to estimate the range information in Phase I with imperfectly synchronized BS and UEs, and how to localize the targets in Phase II via data association and removing UEs with quite erroneous estimated locations. 

\subsection{Phase \uppercase\expandafter{\romannumeral1}: Range Estimation}
\label{sec: phase 1}
In this section, we introduce the proposed range estimation method based on the received signals in Phase \uppercase\expandafter{\romannumeral1} of the two-phase sensing protocol. 
It can be shown that the downlink signal received by the BS in the frequency domain is given by \cite{shi2022}
\begin{equation}
\bar{\boldsymbol{y}} = \left[\bar{y}_1,\dots,\bar{y}_{N_d} \right]^T 
= \sqrt{p_0} \diag{(\boldsymbol{s})}\boldsymbol{G}\boldsymbol{h}^{\rm BTB} + \bar{\boldsymbol{z}},
\end{equation}
where $\diag{(\boldsymbol{s})}$ is a diagonal matrix with the diagonal elements being $\boldsymbol{s}$, $\boldsymbol{G} \in \mathbb{C}^{N_d \times L}$ with the $(n,l)$-th element being $G_{n,l} = e^{\frac{-j2\pi (n-1)(l-1)}{N_d}}$, and $\bar{\boldsymbol{z}} \sim \mathcal{CN}(0,\sigma^2 \boldsymbol{I}_{N_d})$ is the receiver noise at the BS with $\boldsymbol{I}_{N_d}$ being an identity $N_d \times N_d$ matrix.
Since there are only a few targets in the system, $\boldsymbol{h}^{\rm BTB}$ is a sparse vector with a few non-zero elements. Thus, the LASSO technique can be utilized to estimate the time-domain channel $\boldsymbol{h}^{\rm BTB}$ by solving the following problem
\begin{align}\label{pb}
\!\!\!\!\!	\mbox{}\!\!\underset{\scriptstyle \boldsymbol{h}^{\rm BTB}}{\min}\!\! & \quad \frac{1}{2}\|\bar{\boldsymbol{y}}-\sqrt{p_0}\diag(\boldsymbol{s})\boldsymbol{G}\boldsymbol{h}^{\rm BTB}\|_2^2 + \lambda\|\boldsymbol{h}^{\rm BTB}\|_1,
\end{align}
where $\lambda > 0$ is a given coefficient to make sure that the solution to problem \eqref{pb} is sparse. 
Problem \eqref{pb} is convex and can be efficiently solved by using existing solvers, e.g., CVX. 
Let $\bar{\boldsymbol{h}}^{\rm BTB} = \left[\bar{h}_1^{\rm BTB},\dots,\bar{h}_L^{\rm BTB}\right]^T$ denote the optimal solution of problem \eqref{pb}. Note that $\bar{h}_{l}^{\rm BTB} \neq 0$ holds if and only if there is a target $\bar{k}_{l}$ such that the delay of the propagation path from the BS to the target $\bar{k}_{l}$ to the BS is of $l$ downlink OFDM sample period. 
Then, the range between the BS and target $\bar{k}_l$, i.e., $d^{\rm BT}_{\bar{k}_{l}}$, is estimated as follows \cite{shi2022}
\begin{equation}\label{dbt}
\bar{d}^{\rm BT}_{\bar{k}_l} = \frac{lc_0}{2N_d\Delta f_d} + \frac{c_0}{4N_d\Delta f_d},
\end{equation}
where $c_0$ denotes the speed of light. To summarize, the BS will have a set consisting of the estimated ranges from the BS to all the targets, i.e., 
\begin{equation}\label{Dbt}
    \mathcal{D}^{\rm BT} = \{\bar{d}^{\rm BT}_{\bar{k}_l}|\forall l~\text{satisfying}~\bar{h}^{\rm BTB}_l \neq 0\}.
\end{equation}

Next, we focus on estimating the ranges of the paths from the BS to the targets to UE $m$, i.e., $d_{m,k}^{\rm BTU}$'s, based on the received downlink signals at UE $m$, $\forall m$. 
To this end, we first need to estimate $\boldsymbol{h}_{m}^{\rm BU}$'s. However, different from the case for processing the signals received by the BS as shown in the above, the unknown STOs, i.e.,  $\tau_m$'s, make it hard to estimate $h_{m,l}^{\rm BU}$'s based on the signals received by the UEs, i.e., \eqref{bs2ue}.  
To tackle this challenge, we first define 
 \begin{equation}\label{eqn:LOS delay}
     l_m = \left \lfloor \frac{N_d\Delta f_d \|\boldsymbol{b}-\boldsymbol{u}_m\|}{c_0} \right \rfloor, ~ \forall m,
 \end{equation}
as the propagation delay (in terms of downlink OFDM sample periods) of the LOS path from the BS to UE $m$, where $\lfloor \cdot \rfloor$ is the floor function.
Then, we can transform this asynchronous system into a so-called \textit{quasi-synchronous} system by reformulating \eqref{bs2ue} as \cite{Morelli2007}, \cite{Barbarossa2002}
\begin{align}
	y_{m,n}^{\rm DL} & \overset{(a)}{=} \sum_{l=l_m}^{L-1} h_{m,l}^{\rm BU} \bar{\chi}_{n-l-\tau_m} + \check{z}_{m,n} \nonumber \\[-1mm]
	& = \sum_{l=l_{m}+\tau_m}^{L+\tau_m-1} h_{m,l-\tau_m}^{\rm BU} \bar{\chi}_{n-l} + \check{z}_{m,n} \nonumber \\[-1mm]
	& = \sum_{l=0}^{L+\tau_{\max}-1}\tilde{h}_{m,l}^{\rm BU}\bar{\chi}_{n-l} + \check{z}_{m,n}, \forall m, n, \label{virtual}
\end{align}
where
$(a)$ holds as no path with a delay of $l < l_m$ downlink OFDM sample periods exits between the BS and UE $m$, i.e., $h_{m,l}^{\rm BU} = 0, l \in [0, l_m)$, $\tau_{\max} = \max_m |\tau_{m}|$ is defined as the maximum absolute STO between the UEs and the BS, and $\tilde{h}_{m,l}^{\rm BU}$ is defined as
\begin{equation}\label{tildehbu}
\tilde{h}_{m,l}^{\rm BU} = \begin{cases}
&h_{m,l-\tau_m}^{\rm BU}, \quad  \text{if}~l \in [l_m+\tau_m,L+\tau_m-1],\\
&0, \quad \quad \quad \quad \text{otherwise}.
\end{cases}
\end{equation}  
Therefore, $\tilde{h}_{m,l}^{\rm BU}$ can be interpreted as the extended channel associated with a path between the BS and UE $m$ \cite{Morelli2007}, where the imperfectly synchronized UE $m$ believes the path delay to be $l$ downlink OFDM sample period, but it is actually $l-\tau_m$ downlink OFDM sample periods. In this paper, we assume that $l_m + \tau_m \geq 0$, such that UE $m$ sees no inter-symbol interference (ISI) from the next OFDM symbol sent by the BS even when the BS's clock is earlier than that at UE $m$, i.e., $\tau_m < 0$, $\forall m$. 

Note that there is no STO, i.e., $\tau_m$, in the reformulated signal model \eqref{virtual}, implying that $\tilde{h}_{m,l}^{\rm BU}$'s can be estimated by applying the conventional OFDM channel estimation techniques. However, the propagation delays estimated based on $\tilde{h}_{m,l}^{\rm BU}$ will be shifted by the unknown STO, i.e., $\tilde{h}_{m,l}^{\rm BU} \neq 0$ indicates that there is a path from the BS to UE $m$ whose propagation delay is of $l-\tau_m$, instead of $l$, downlink OFDM sample periods. Thus, we can never obtain correct delay/range estimations without knowing the STOs. In the following, we propose an efficient method that can first estimate the STOs based on the LOS signals from the BS to the UEs, and then utilize the STOs to estimate the propagation delays from the BS to the target to the UEs. 
 
Specifically, according to \eqref{virtual}, the received downlink signal at UE $m$ in the frequency domain is given by 
\begin{align}
	\bar{\boldsymbol{y}}_m^{\rm DL} &= \left[\bar{y}_{m,1}^{\rm DL},\dots,\bar{y}_{m,N_d}^{\rm DL} \right]^T \nonumber \\[-1mm]
	&= \sqrt{p_0} \diag{(\boldsymbol{s})}\boldsymbol{G}_m\tilde{\boldsymbol{h}}^{\rm BU}_m + \bar{\boldsymbol{z}}_m^{\rm DL}, \forall m,
\end{align}
where $\boldsymbol{G}_m \in \mathbb{C}^{N_d \times (L+\tau_{\max})}$ with the $(n,l)$-th element being $G_{m,n,l} = e^{\frac{-j2\pi (n-1)(l-1)}{N_d}}$, $\tilde{\boldsymbol{h}}^{\rm BU}_m = [\tilde{h}_{m,0}^{\rm BU},\dots,\tilde{h}_{m,L+\tau_{\max}-1}^{\rm BU}]$, and $\bar{\boldsymbol{z}}_m^{\rm DL} \sim \mathcal{CN}(0,\check{\sigma}_m^2 \boldsymbol{I}_{N_d})$ denotes the noise at UE $m$.

In this paper, we assume that all the UEs know the signal $\boldsymbol{s}$ sent by the BS (e.g., $\boldsymbol{s}$ can be the known pilot signal). Moreover, $\tilde{\boldsymbol{h}}_m^{\rm BU}$ is a sparse channel vector. Thus, the extended channel between the BS and UE $m$ as defined in \eqref{tildehbu} can be estimated by solving the following LASSO problem
\begin{align}\label{pu}
\!\!\!\!\!	\mbox{}\!\!\underset{\scriptstyle \tilde{\boldsymbol{h}}_m^{\rm BU}}{\min}\!\! & \quad \frac{1}{2}\|\bar{\boldsymbol{y}}_m^{\rm DL}-\sqrt{p_0}\diag(\boldsymbol{s})\boldsymbol{G}_m\tilde{\boldsymbol{h}}^{\rm BU}_m\|_2^2 + \lambda\|\tilde{\boldsymbol{h}}^{\rm BU}_m\|_1.
\end{align}
Let $\bar{\boldsymbol{h}}_m^{\rm BU} = [\bar{h}_{m,0}^{\rm BU},\dots,\bar{h}_{m,L_m-1}^{\rm BU}]^T$ denote the optimal solution to problem \eqref{pu}. If $\bar{h}_{m,l}^{\rm BU} \neq 0$ for some $l$, then there is a path between the BS and UE $m$, whose propagation delay is of $l- \tau_m$ downlink OFDM sample periods. Thus, we can define 
\begin{equation}\label{eqn:estiamted LOS}
    \bar{l}_m = \min \{l| \forall l~\text{with}~\bar{h}^{\rm BU}_{m,l} \neq 0\}
\end{equation}
as the estimated propagation delay of the direct path between the BS and UE $m$, when their clocks differ by $\tau_m$ downlink OFDM sample periods, $\forall m$. 
Note that the true propagation delay of the direct path between the BS and UE $m$ should be $l_m$ given in (\ref{eqn:LOS delay}). However, in practice, we merely know the erroneous positions of the UEs. Define the propagation delay approximated by (\ref{eqn:LOS delay}) as
\begin{equation}\label{lm}
\tilde{l}_m = \left \lfloor \frac{N_d \Delta f_d \|\boldsymbol{b}-\hat{\boldsymbol{u}}_m\|}{c_0} \right \rfloor, \forall m,
\end{equation}where the true positions of the UEs are replaced by their estimated positions. 
Then, the STO between the BS and UE $m$ is estimated as 
\begin{equation}\label{taum}
    \hat{\tau}_m = \bar{l}_m - \tilde{l}_m, \forall m.
\end{equation}
Next, for the other $l$'s ($l \neq \bar{l}_m$), $\bar{h}^{\rm BU}_{m,l} \neq 0$ holds if and only if there is a target $\bar{k}_{m,l}$ such that the propagation delay of the path from the BS to target $\bar{k}_{m,l}$ to UE $m$ is of $l$ downlink OFDM sample period. Thus, the range from the BS to target $\bar{k}_{m,l}$ to UE $m$ is estimated as
\begin{equation}\label{dbtu}
    \bar{d}_{m,\bar{k}_{m,l}}^{\rm BTU} = \frac{(l-\hat{\tau}_m)c_0}{N_d\Delta f_d} + \frac{c_0}{2N_d \Delta f_d}, \forall m.
\end{equation}

To summarize, following Phase \uppercase\expandafter{\romannumeral1}, UE $m$ will obtain a set of ranges from the BS to all the targets to UE $m$, i.e., 
\begin{equation}\label{Dbtu}
    \mathcal{D}^{\rm BTU}_{m} = \{\bar{d}_{m,\bar{k}_{m,l}}^{\rm BTU}|\forall l~\text{satisfying}~\bar{h}^{\rm BU}_{m,l}\neq 0~\text{and}~l \neq \bar{l}_m\}. 
\end{equation}

\subsection{Phase II: Target Localization via Joint Data Association and Ineffective Anchors Removing}\label{sec:Phase II: Target Localization via Joint Data Association and Ineffective Anchors Removing}

In Phase \uppercase\expandafter{\romannumeral2}, we aim to localize the targets with the knowledge about $\mathcal{D}^{\rm BTU}_m$'s and $\mathcal{D}^{\rm BT}$ based on the multilateration method \cite{Torrieri1984}. 
To successfully localize the targets, it requires both accurate information about the anchor positions and the ranges from the targets to the anchors. However, in practice, the positions of some UEs estimated via GPS may be highly inaccurate.
Moreover, if $\hat{\boldsymbol{u}}_m$ is quite inaccurate compared to $\boldsymbol{u}_m$, then $\tilde{l}_m$ given in \eqref{lm} is a poor estimation of the propagation delay for the LOS path between the BS and UE $m$, leading to wrong estimation of the STO $\tau_m$ given in \eqref{taum} as well as these of $\bar{d}^{\rm BTU}_{m,\bar{k}_{m,l}}$'s given in \eqref{dbtu}. 
This motivates us to remove all the \emph{ineffective} UEs, which are defined as the UEs whose locations estimated by GPS are very inaccurate, from the anchor set, and merely utilize the \emph{effective} UEs, that are defined as the UEs whose locations estimated by GPS are quite accurate, as the anchors to localize the targets. 
In addition, given the effective UEs selected as anchors, we also need to find the data association solution, i.e., for each target $k$, which elements in $\mathcal{D}^{\rm BT}$ and $\mathcal{D}^{\rm BTU}_m$ are $\bar{d}_{k}^{\rm BT}$ and $\bar{d}_{m,k}^{\rm BTU}$, respectively, 
such that we can apply the multilateration method to localize the targets. 
\subsubsection{Problem Formulation for Joint Data Association and Ineffective UE Removing}
First, define $\tilde{\mathcal{M}}$ as the set of effective UEs that are used as the anchors. Therefore, for any UE $m \notin \tilde{\mathcal{M}}$, its ranges set $\mathcal{D}_m^{\rm BTU}$ is not used to localize the target. 
Moreover, define $g_{0,k}$ such that the estimated range for the path from the BS to target $k$, i.e., $\bar{d}_{k}^{\rm BT}$ shown in \eqref{dbt}, is the $g_{0,k}$-th largest element in $\mathcal{D}^{\rm BT}$, i.e., $\bar{d}_{k}^{\rm BT} = \mathcal{D}^{\rm BT}(g_{0,k})$, $\forall k$, where $\mathcal{B}(b)$ denotes the $b$-th largest element in $\mathcal{B}$.
To mitigate the ambiguity in target indexing, we define target $k$ as the target whose range to the BS is the $k$-th largest element in $\mathcal{D}^{\rm BT}$, i.e.,  
\begin{equation}\label{g0k}
	g_{0,k} = k, \forall k. 
\end{equation}
 
Moreover, given any $m \in \tilde{\mathcal{M}}$, define $g_{m,k}$ such that the estimated range for the path from the BS to target $k$ to effective UE $m$, i.e., $\bar{d}_{m,k}^{\rm BTU}$ shown in \eqref{dbtu}, is the $g_{m,k}$-th largest element in $\mathcal{D}^{\rm BTU}_m$, $\forall k$, i.e., $\bar{d}_{m,k}^{\rm BTU} = \mathcal{D}^{\rm BTU}_m(g_{m,k})$. For convenience, we further define $\mathcal{G}_{k} = \{g_{m,k}|\forall m \in \tilde{\mathcal{M}}\}$
as the data association solution for target $k$ to the effective UEs in $\tilde{\mathcal{M}}$, $\forall k$. Because each UE $m$ has a range set $\mathcal{D}_m^{\rm BTU}$ consisting of $K$ elements, and any two elements in $\mathcal{D}_m^{\rm BTU}$ belong to two targets, the data association solution should satisfy 
\begin{align}
	g_{m,k} \in \{1,\dots,K\}, \forall k, \forall m \in  \tilde{\mathcal{M}}, \label{c1} \\
	g_{m,k} \neq g_{m,\bar{k}}, ~\forall k \neq \bar{k}, m \in  \tilde{\mathcal{M}}.\label{c2} 
\end{align}

Given any effective UE set $\tilde{\mathcal{M}}$ and data association solution $\mathcal{G}_k$, any target $k$ can be localized based on the multilateration method by solving the following weighted nonlinear least squares problem \cite{caffery1998}
\begin{align*} 
\!\!\!\!\!	\mbox{(P1-$k$): }\underset{\scriptstyle \boldsymbol{a}_k}{\min}\!\! & \quad 
  v f_0(\boldsymbol{a}_k) + \sum_{m \in \tilde{\mathcal{M}}} f_m(\boldsymbol{a}_k,\mathcal{G}_k), \label{p1}
\end{align*}
where 
\vspace{-1mm} 
\begin{align}
	f_0(\boldsymbol{a}_k) 
	&\!=\! \left(\|\boldsymbol{b}-\boldsymbol{a}_k\|\!-\! \mathcal{D}^{\rm BT}(k)\right)^2, \forall k
\end{align}
\vspace{-1mm}
and
\vspace{-1mm}
\begin{align}	
	f_m(\boldsymbol{a}_k,\mathcal{G}_k)  
	&\!=\!  \big(\|\hat{\boldsymbol{u}}_m-\boldsymbol{a}_k\|\!-\! \mathcal{D}^{\rm BTU}_m(g_{m,k})\!+\!\mathcal{D}^{\rm BT}(k)\big)^2 \forall m, k
\end{align}
denote the residuals for localizing target $k$ associated with the BS and UE $m$, respectively, and $v$ is the weight of the residual corresponding to the BS.
 Note that the residual associated with the BS is much more trustworthy than that associated with a UE, because the estimated positions of the UEs are subject to errors. 
Thus, we set $v > 1$, such that the residual associated with the BS plays a more significant role in determining the location of the target. 
 
Given any effective UE set $\tilde{\mathcal{M}}$ and data association solution $\mathcal{G}_k$, we can apply the Gauss-Newton method \cite{Torrieri1984} to solve problem (P1-$k$) to localize target $k$, $\forall k$.  
Define $\theta_k (\tilde{\mathcal{M}}, \mathcal{G}_{k})$ as the objective value of problem (P1-$k$) achieved by the Gauss-Newton method for localizing target $k$ given $\tilde{\mathcal{M}}$ and $\mathcal{G}_{k}$, $\forall k$.  
To reduce the dependence on the number of anchors for localization, we further define
\begin{equation}
\bar{\theta}_k(\tilde{\mathcal{M}},\mathcal{G}_{k}) = \frac{\theta_k (\tilde{\mathcal{M}},\mathcal{G}_{k})}{|\tilde{\mathcal{M}}|+1} \label{residual}
\end{equation}
as the corresponding normalized residual associated with target $k$ \cite{chen1999}, $\forall k$. 
Intuitively, given the right effective UE set and data association solution, the true locations of the targets can lead to very small estimation residue. On the other hand, the wrong effective UE set and data association solution can lead to very wrong target location estimation and large estimation residue. 
Thus, we aim to jointly estimate $\tilde{\mathcal{M}}$ and $ \{ \mathcal{G}_{1},\dots,\mathcal{G}_{K}\}$ by solving the following problem
\begin{equation*}
\begin{aligned}
\mbox{(P2): } \min_{\tilde{\mathcal{M}},\{\mathcal{G}_{1},\dots,\mathcal{G}_{K}\}} \quad & \sum_{k=1}^{K} \bar{\theta}_k(\tilde{\mathcal{M}},\mathcal{G}_{k})\\
\textrm{s.t.} \quad \quad \quad \!\! &  \eqref{c1}, \eqref{c2}.
\end{aligned}
\end{equation*} 

One straightforward way to solve the above problem is via exhaustive search. Specifically, all the effective UE set and data association solutions that satisfy \eqref{c1} and \eqref{c2} should be listed, and the one that minimizes problem (P2) is set as the optimal effective UE set and data association solution. 
However, this method is of prohibitively high complexity as we need to solve problems (P1-$k$), $\forall k$, many times. This motivates us to propose a low-complexity algorithm for solving problem (P2). In the following, we first consider a simplified case of problem (P2) when there is only one target, i.e., $K=1$. 
In this case, data association is no longer an issue, and the challenge for target localization is just the effective UE set selection. 
Then, based on the effective UE selection method under the single-target case, we propose an efficient joint effective UE selection and data association method for the multi-target case.  

\subsubsection{Single-Target Localization via Ineffective UE Removal}
Consider the case when there is only one target in the network, i.e., $K = 1$. We refer to this target as target 1. In this case, there is only one element in $\mathcal{D}^{\rm BT}$ and $\mathcal{D}^{\rm BTU}_m$, respectively, $\forall m$. 
Thus, the data association variables are
\begin{equation}\label{single_da}
g_{m,1} = 1, \forall m \in \mathcal{M}.  
\end{equation} 
Given the above data association solution, problem (P2) reduces to the following effective UE selection problem 
\begin{equation*}
\begin{aligned}
\mbox{(P3): } \min_{\tilde{\mathcal{M}}} \quad &  \bar{\theta}_1(\tilde{\mathcal{M}})
\end{aligned}
\end{equation*}

In the following, we propose an iterative algorithm with low complexity to solve problem (P3).
Specifically, we adopt the outlier detection technique \cite{li2006, Guvenc2009} to remove the ineffective UEs with inaccurate position information from GPS such that the BS together with the effective UEs with accurate position information from GPS can jointly serve as the anchors to localize the target.
Our proposed iterative algorithm starts with the original set $\mathcal{M}$, and removes one ineffective UE with inaccurate position estimated by GPS in each iteration, until no significant gain is obtained by removing an UE. 
Specifically, define $\tilde{\mathcal{M}}_i$ as the set of UEs that are not removed from $\mathcal{M}$ after the $i$-th iteration with $|\tilde{\mathcal{M}}_i| = M-i$, $\forall i$. 
Note that in the initialization step, we set $\tilde{\mathcal{M}}_0 = \mathcal{M}$. The way to remove one UE from set $\tilde{\mathcal{M}}_{i-1}$ in the $i$-th iteration of our proposed algorithm is as follows. 
Define 
\begin{equation}
    \mathcal{Q}_i = \{\mathcal{M}_{i}|\mathcal{M}_{i} \subseteq \tilde{\mathcal{M}}_{i-1}, |\mathcal{M}_{i}| = |\tilde{\mathcal{M}}_{i-1}|-1\}
\end{equation}
as the set of all possible solutions in the $i$-th iteration of the algorithm. According to problem (P3), at the $i$-th iteration of our algorithm, we set $\tilde{\mathcal{M}}_i$ as the optimal solution to the following problem
\begin{equation*}
\begin{aligned}
\mbox{(P4): } \min_{{\mathcal{M}}_i} \quad &  \bar{\theta}_1({\mathcal{M}}_i)\\
\textrm{s.t.} \quad & {\mathcal{M}}_i \in \mathcal{Q}_i. 
\end{aligned}
\end{equation*}

Note that problem (P4) can be solved via the exhaustive research method efficiently because $|\mathcal{Q}_i| = M-i+1, \forall i$. Therefore, in the $i$-th iteration of the algorithm, if the removal of a UE from the anchor set $\tilde{\mathcal{M}}_{i-1}$ can lead to the minimum normalized residual for localizing the target, we just remove this UE. If after the $i^\ast$-th iteration,
\begin{equation}\label{stopcriterion}
    |\bar{\theta}_1(\tilde{\mathcal{M}}_{i^\ast})-\bar{\theta}_1(\tilde{\mathcal{M}}_{i^\ast-1})| \leq \theta_{\rm th},
\end{equation}
 where $\theta_{\rm th}$ is a given threshold, then it indicates that removing a UE from the anchor set can no longer significantly reduce the localization residual. Then, we will terminate our algorithm. Next, the set $\tilde{\mathcal{M}}_{i^\ast}$ serves as the solution to problem (P3), which is denoted by $\tilde{\mathcal{M}}^\ast$. 
  Last, given $\tilde{\mathcal{M}}^\ast$, the solution of (P1-1) is the final estimation of the target location. The above procedure for solving problem (P3) is summarized in Algorithm~\ref{alg1}. 

This algorithm is of low complexity. Specifically, with the exhaustive search method for problem (P3), we need to solve problem (P1) for 
$D_1= \sum_{n=2}^M \tbinom{M}{n} = \sum_{n=2}^{M}\frac{M!}{n!(M-n)!}$ 
times. However, under our proposed algorithm, in the $i$-th iteration ($i \geq 1$), we only need to solve problem (P1) for $M-i+1$ times. The algorithm reaches its worst-case complexity when it stops at $i = M - 2$ (when only two UEs and the BS serve as three anchors). In this case, we need to solve problem (P1) for 
$D_2 = \sum_{i=1}^{M-2}M-i+1 = \frac{(M-2)(M+3)}{2}$ times.

\begin{algorithm}[t]
	{\bf Input}: $\mathcal{M}$, $\mathcal{D}^{\rm BT}$, $\mathcal{D}_m^{\rm BTU}$, $\hat{\boldsymbol{u}}_m$, $\boldsymbol{b}$, $g_{m,1} = 1, \forall m \in \mathcal{M}$.  \\
	{\bf Initialization}: Set $ \tilde{\mathcal{M}}_0 = \mathcal{M}$ and $i = 1$. Given $\tilde{\mathcal{M}}_0$, solve problem (P1-1) to obtain $\bar{\theta}_1(\tilde{\mathcal{M}}_0)$.\\
	{\bf Repeat:}
		\begin{enumerate}
			\item[1.] Given any $\mathcal{M}_i \in \mathcal{Q}_i$, solve problem (P1-1) to obtain $\bar{\theta}_1(\mathcal{M}_i)$ and the corresponding estimation of the target's location; 
			\item[2.] Set $\tilde{\mathcal{M}}_i$ and $\bar{\theta}_1(\tilde{\mathcal{M}}_i)$ as the optimal solutions and the optimal value of problem (P4);
			\item[3.] Set $i = i + 1$;
		\end{enumerate}
	{\bf Until} $|\bar{\theta}_1(\tilde{\mathcal{M}}_{i}) - \bar{\theta}_1(\tilde{\mathcal{M}}_{i-1})| \leq \theta_{\rm th}$.\\
	{\bf Output}: The effective UE set $\tilde{\mathcal{M}}^\ast$ and the corresponding estimation of target location $\boldsymbol{a}_1^\ast$.	
	\caption{Algorithm to Solve Problem (P3) for Single-Target Localization via Ineffective UE Removing}
	\label{alg1}
\end{algorithm}

\subsubsection{Multi-Target Localization via Joint Data Association and Ineffective Anchors Removal}
Next, we consider the general multi-target scenario. In this case, besides the effective UE set $\tilde{\mathcal{M}}$, we also need to know the data association solution, i.e., $\mathcal{G}_1, \dots, \mathcal{G}_K$, to localize all the targets. Note that given any feasible data association solution that satisfies \eqref{c1} and \eqref{c2}, the distances from each target to all the UEs are given, and we can apply Algorithm~\ref{alg1} to localize each target independently. Then, we can apply the exhaustive search method to select the data association solution that leads to the minimum localization residue as the optimal data association solution to problem (P2), and set the corresponding estimated locations as the localization results. However, the above approach does not utilize a property: the effective and ineffective UE sets are common to all the targets, i.e., if a UE has accurate/inaccurate estimated location via GPS, it is useful/not useful to localize all the targets. This indicates that we do not need to determine the effective UE set every time when we aim to localize a target. Instead, if the accurate effective UE set can be known after localizing a target, we should just use this set and only consider data association when localizing the other targets. Based on this idea, in the following, we aim to design a low-complexity UE selection and data association algorithm to localize multiple targets, based on the UE selection method introduced in Algorithm \ref{alg1}. 

{\bf Step 1: Using Target 1 to Obtain Effective UE Set and Target 2 to Check its Accuracy} 

For convenience, define $\tilde{\mathcal{G}}_k = \{g_{m,k}|\forall m \in \mathcal{M}\}$ as the data association solution for target $k$ to all the UEs in $\mathcal{M}$, $\forall k$.  
Then, we can define a set that consists of all the feasible data association solutions for target 1 to all the UEs, which is 
\begin{equation}\label{h1}
	{\mathcal{H}}_1 = \left \{\tilde{\mathcal{G}}_1|g_{m,1} \in \mathcal{K}, \forall m \in \mathcal{M} \right \}.
\end{equation} 
Given each feasible data association solution of target 1 $\tilde{\mathcal{G}}_1 \in {\mathcal{H}}_1$, we can use Algorithm~\ref{alg1} to localize target 1. Given this data association solution, let $\boldsymbol{a}_1(\tilde{\mathcal{G}}_1)$, $\tilde{\mathcal{M}}(\tilde{\mathcal{G}}_1)$, and $\bar{\theta}_1(\tilde{\mathcal{M}}(\tilde{\mathcal{G}}_1),\tilde{\mathcal{G}}_1)$ denote the estimated location of target 1, the effective UE set, and the estimation residue obtained by Algorithm~1. Define $\tilde{\mathcal{G}}_1^\ast = \{g_{m,1}^\ast|\forall m \in \mathcal{M}\}$ as the optimal solution to the following problem 
\begin{equation*}
	\begin{aligned}
	 {\mbox{(P5-1-M): }} \min_{\tilde{\mathcal{G}}_1} \quad &  \bar{\theta}_1(\tilde{\mathcal{M}}(\tilde{\mathcal{G}}_1),\tilde{\mathcal{G}}_1)\\
		\textrm{s.t.} \quad & \tilde{\mathcal{G}}_1 \in \mathcal{H}_1, 
	\end{aligned}
\end{equation*}
 which can be obtained via the exhaustive search method. At last, the estimated location of target 1, the effective UE set, and the corresponding residue are set as $\boldsymbol{a}_1(\tilde{\mathcal{G}}_1^\ast)$, $\tilde{\mathcal{M}}(\tilde{\mathcal{G}}_1^\ast)$, and $\bar{\theta}_1(\tilde{\mathcal{M}}(\tilde{\mathcal{G}}_1^\ast),\tilde{\mathcal{G}}_1^\ast)$, respectively.

After the effective UE set $\tilde{\mathcal{M}}(\tilde{\mathcal{G}}_1^\ast)$ is obtained when localizing target 1, one straightforward approach is to fix this set when localizing the other $K-1$ targets. In other words, we merely perform data association to localize these targets. However, if the effective UE set $\tilde{\mathcal{M}}(\tilde{\mathcal{G}}_1^\ast)$ is wrong, then this error will be propagated to the localization of the other targets. To tackle this issue, the following approach is proposed. Specifically, we merely use $\tilde{\mathcal{M}}(\tilde{\mathcal{G}}_1^\ast)$ as the effective UE set to localize target 2. Note that besides \eqref{c1}, the data association solution of target 2 should also be different from that of target 1 according to \eqref{c2}, i.e., 
\begin{equation}\label{uni12}
	g_{m,2} \neq g_{m,1}^\ast, \forall m \in \mathcal{M}. 
\end{equation} 
Therefore, we can define the set consisting of all the feasible data association solutions of target 2 to the UEs in $\tilde{\mathcal{M}}(\tilde{\mathcal{G}}_1^\ast)$ as 
\begin{equation}\label{H2}
	\mathcal{H}_2 = \left\{{\mathcal{G}}_2|g_{m,2} \neq g_{m,1}^\ast, g_{m,2} \in \mathcal{K},  \forall m \in \tilde{\mathcal{M}}(\tilde{\mathcal{G}}_k^\ast)\right\}.
\end{equation}
Given $\tilde{\mathcal{M}}(\tilde{\mathcal{G}}_1^\ast)$ and any ${\mathcal{G}}_2 \in \mathcal{H}_2$, let $\bar{\theta}_2 (\tilde{\mathcal{M}}(\tilde{\mathcal{G}}_1^\ast),\mathcal{G}_2)$ denote the estimation residue which can be obtained by solving problem (P1-2) via the Gauss-Newton method. Then, given $\tilde{\mathcal{M}}(\tilde{\mathcal{G}}_1^\ast)$, the optimal data association solution of target 2, which is denoted as $\mathcal{G}_2^\ast$, can be obtained by solving the following problem 
\begin{equation*}
	\begin{aligned}
	{\mbox{(P5-1-C): }}	\min_{{\mathcal{G}}_2} \quad &  \bar{\theta}_2 (\tilde{\mathcal{M}}(\tilde{\mathcal{G}}_1^\ast),\mathcal{G}_2)\\
		\textrm{s.t.} \quad & {\mathcal{G}}_2 \in \mathcal{H}_2.
	\end{aligned}
\end{equation*}
The corresponding estimation residue is denoted as $\bar{\theta}_2 (\tilde{\mathcal{M}}(\tilde{\mathcal{G}}_1^\ast),\mathcal{G}_2^\ast)$. Let $\bar{\theta}_{\rm th}$ denote a pre-designed threshold. Then, if $\bar{\theta}_2 (\tilde{\mathcal{M}}(\tilde{\mathcal{G}}_1^\ast),\mathcal{G}_2^\ast) \leq \bar{\theta}_{\rm th}$, this indicates that the effective UE set $\tilde{\mathcal{M}}(\tilde{\mathcal{G}}_1^\ast)$ leads to small estimation residue when localizing both targets 1 and 2, Therefore, we can trust $\tilde{\mathcal{M}}(\tilde{\mathcal{G}}_1^\ast)$ to be the right effective UE set solution. Otherwise, if $\bar{\theta}_2 (\tilde{\mathcal{M}}(\tilde{\mathcal{G}}_1^\ast), \mathcal{G}_2^\ast) > \bar{\theta}_{\rm th}$, then, the effective UE set $\tilde{\mathcal{M}}(\tilde{\mathcal{G}}_1^\ast)$ may not be the right solution because it only leads to small residue to localize target 1, but leads to large residue to localize target 2. In this case, we need to check whether we can obtain the correct effective UE set via target 2 in Step 2. 

{\bf Step 2 (when Step 1 fails to find the effective UE set): Using Target 2 to Obtain Effective UE Set and Target 3 to Check its Accuracy} 

Step 2 is similar to Step 1. Instead of target 1, we use target 2 to obtain the effective UE set, denoted by $\tilde{\mathcal{M}}(\tilde{\mathcal{G}}_2^\ast)$, where $\tilde{\mathcal{G}}_2^\ast$ is the optimal data association solution of target 2 to all the UEs in $\mathcal{M}$. Then, we can use $\tilde{\mathcal{M}}(\tilde{\mathcal{G}}_2^\ast)$ to localize target 3. Let $\bar{\theta}_3(\tilde{\mathcal{M}}(\tilde{\mathcal{G}}_2^\ast),\mathcal{G}_3^\ast)$ denote the estimation residue given $\tilde{\mathcal{M}}(\tilde{\mathcal{G}}_2^\ast)$ and the optimal data association solution $\mathcal{G}_3^\ast$. If $\bar{\theta}_3(\tilde{\mathcal{M}}(\tilde{\mathcal{G}}_2^\ast),\mathcal{G}_3^\ast) \leq \bar{\theta}_{\rm th}$, then we claim that $\tilde{\mathcal{M}}(\tilde{\mathcal{G}}_2^\ast)$ is the right effective UE set. Otherwise, if $\bar{\theta}_3(\tilde{\mathcal{M}}(\tilde{\mathcal{G}}_2^\ast),\mathcal{G}_3^\ast) > \bar{\theta}_{\rm th}$, we need to check whether we can obtain the effective UE set via target 3 in Step 3. 

If the estimation residue is larger than the threshold at all the previous steps, we can perform the above process in the next step. This process will not end until at Step $k^\ast$, where 
\begin{equation}
	\bar{\theta}_{k^\ast+1}(\tilde{\mathcal{M}}(\tilde{\mathcal{G}}_{k^\ast}^\ast),\mathcal{G}_{k^\ast+1}^\ast) \leq \bar{\theta}_{\rm th},
\end{equation}
holds for the first time. Then, we claim that $\tilde{\mathcal{M}}(\tilde{\mathcal{G}}_{k^\ast}^\ast)$ is the effective UE set solution. Moreover, target $k^\ast$ and target $k^\ast + 1$ have been localized in Step $k^\ast$. Let $\boldsymbol{a}_{k^\ast}^\ast$ and $\boldsymbol{a}_{k^\ast+1}^\ast$ denote their estimated locations, and $\mathcal{G}_{k^\ast}^\ast = \{g_{m,k^\ast}^\ast|m \in \tilde{\mathcal{M}}(\tilde{\mathcal{G}}_{k^\ast}^\ast)\}$ and $\mathcal{G}_{k^\ast+1}^\ast = \{g_{m,k^\ast+1}^\ast|m \in \tilde{\mathcal{M}}(\tilde{\mathcal{G}}_{k^\ast}^\ast)\}$ denote their data association solutions to the effective UEs in $\tilde{\mathcal{M}}(\tilde{\mathcal{G}}_{k^\ast}^\ast)$, respectively.  

{\bf Last Step: Localizing Other Targets Given Effective UE Set $\tilde{\mathcal{M}}(\mathcal{G}_{k^\ast}^\ast)$}

Define the set of targets that have not been localized as $\bar{\mathcal{K}} = \{\forall k \in \mathcal{K}, k \neq k^\ast, k \neq k^\ast+1\}$. Then, given the effective UE set $\tilde{\mathcal{M}}(\mathcal{G}_{k^\ast}^\ast)$ and the data association solutions of target $k^\ast$ and $k^\ast+1$, i.e., $\mathcal{G}_{k^\ast}^\ast$ and $\mathcal{G}_{k^\ast+1}^\ast$, the joint data association and localization problem for the other $K-2$ targets are
\begin{equation}
	\begin{aligned}
		\mbox{(P6): } \min_{\{\mathcal{G}_k| k \in \bar{\mathcal{K}}\}} \quad  & \sum_{k \in \bar{\mathcal{K}}} \bar{\theta}_k(\tilde{\mathcal{M}}(\mathcal{G}_{k^\ast}^\ast),\mathcal{G}_k)\\
		\textrm{s.t.} \quad \quad  & g_{m,k} \neq g_{m,k^\ast}^\ast, \forall k \in \bar{\mathcal{K}}, m \in \tilde{\mathcal{M}}(\mathcal{G}_{k^\ast}^\ast),\\
		 & g_{m,k} \neq g_{m,k^\ast+1}^\ast, \forall k \in \bar{\mathcal{K}}, m \in \tilde{\mathcal{M}}(\mathcal{G}_{k^\ast}^\ast),\\
		& \eqref{c1}, \eqref{c2}.
	\end{aligned}
\end{equation} 
The above problem has been solved by Algorithm~2 in \cite{shi2022}. Let $\boldsymbol{a}_k^\ast$ denote the estimated location of target $k$ with $k \neq k^\ast$ and $k \neq k^\ast + 1$. 

The above joint effective UE selection and data association approach to localize the $K$ targets is given in Algorithm \ref{alg2}. 

\begin{algorithm}[t]
	{\bf Input}: $\boldsymbol{b}$, $\mathcal{M}$, $\mathcal{K}$, $\mathcal{D}^{\rm BT}$, $\mathcal{D}_m^{\rm BTU}$, $\hat{\boldsymbol{u}}_m, \forall m \in \mathcal{M}$;\\
	{\bf Initialization}: Set $k = 1$;\\
	{\bf Repeat} 
	\begin{enumerate}
		\item[1.] Obtain $\mathcal{H}_k=\{\tilde{\mathcal{G}}_k|g_{m,k} \in \mathcal{K}, \forall m \in \mathcal{M}\}$;
		\item[2.] Given each $\tilde{\mathcal{G}}_k \in \tilde{\mathcal{H}}_k$, localize target $k$ via Algorithm~\ref{alg1} to obtain $\boldsymbol{a}_k(\tilde{\mathcal{G}}_k)$, $\tilde{\mathcal{M}}(\tilde{\mathcal{G}}_k)$, and $\bar{\theta}_k(\tilde{\mathcal{M}}(\tilde{\mathcal{G}}_k),\tilde{\mathcal{G}}_k)$; 
		\item[3.] Set $\tilde{\mathcal{G}}_k^\ast=\{g_{m,k}^\ast|\forall m \in \tilde{\mathcal{M}}(\tilde{\mathcal{G}}_k^\ast)\}$ and $\bar{\theta}_k(\tilde{\mathcal{M}}(\tilde{\mathcal{G}}_k^\ast),\tilde{\mathcal{G}}_k^\ast)$ as the optimal solution and objective value to problem (P5-$k$-M), respectively;    
		\item[4.] Obtain ${\mathcal{H}}_{k+1}=\{{\mathcal{G}}_{k+1}|g_{m,k+1}\neq g_{m,k}^\ast, g_{m,k+1} \in \mathcal{K}, \forall m \in \tilde{\mathcal{M}}(\tilde{\mathcal{G}}_k^\ast)\}$;
		\item[5.] Given $\tilde{\mathcal{M}}(\tilde{\mathcal{G}}_k^\ast)$ and each $\mathcal{G}_{k+1} \in \mathcal{H}_{k+1}$, solve problem (P1-$k$+$1$) to obtain  $\bar{\theta}_{k+1}(\tilde{\mathcal{M}}(\tilde{\mathcal{G}}_k^\ast),\mathcal{G}_{k+1})$ and the corresponding location estimation of target $k+1$;
		\item[6.] Set $\mathcal{G}_{k+1}^\ast$ and $\bar{\theta}_{k+1}(\tilde{\mathcal{M}}(\tilde{\mathcal{G}}_k^\ast),\mathcal{G}_{k+1}^\ast)$ as the optimal solution and objective value to problem (P5-$k$-C), respectively; 
		\item[7.] Set $k=k + 1$;
	\end{enumerate}
	{\bf Until} $\bar{\theta}_{k+1}(\tilde{\mathcal{M}}(\tilde{\mathcal{G}}_k^\ast),\mathcal{G}_{k+1}^\ast) \leq \bar{\theta}_{\rm th}$;
	\begin{enumerate}
		\item[8.] Solve problem (P6) via the method in \cite{shi2022};
	\end{enumerate}
	{\bf Output}: The effective UE set $\tilde{\mathcal{M}}(\mathcal{G}_{k^\ast}^\ast)$ and the corresponding estimations of targets' locations, i.e., $\boldsymbol{a}_1^\ast,\dots,\boldsymbol{a}_K^\ast$.
	
	\caption{Algorithm to Solve Problem (P2) for Multi-Target Localization via Data Association and Ineffective UE Removing}
	\label{alg2}
\end{algorithm}

\section{Localization Method under Active UE based Sensing Mode}\label{sec:Localization Method under Active UE based Sensing Mode}
In this section, we show how to enable UE assisted networked sensing when the UEs can actively send OFDM signals to probe the environment as well. As illustrated in Section \ref{sec:Active UE based Sensing Mode}, under the active UE based sensing mode, each UE $m$ can transmit uplink OFDM signals at it assigned sub-carriers $\mathcal{N}_u^{(m)}$ to probe the environment as well. Therefore, the UEs receive not only the echo signals at the downlink frequency band, i.e., (\ref{bs2ue}), but also the echo signals at the uplink frequency band, i.e., (\ref{u_signal}). In the following, we show how to utilize the additional echo signals at the uplink frequency band to improve the two-phase localization protocol proposed for the passive UE based sensing mode. 

In Phase I, the echo signals received by the UEs at the uplink band, i.e., (\ref{u_signal}), can be used to estimate the distances between the UEs and the targets, i.e., $d_{m,k}^{\rm UT}$'s. Specifically, define
\begin{align}
l_{m,u}=\left \lfloor \frac{N_u \Delta f_u \|\boldsymbol{u}_m - \boldsymbol{u}_u\|}{c_0} \right \rfloor
\end{align}
as the propagation delay (in terms of uplink OFDM sample period) of the LOS path from UE $m$ to UE $u$. 

Then, similar to (\ref{virtual}) in the downlink, the asynchronous signal given in (\ref{u_signal}) can be re-formulated as a quasi-synchronous signal 
\begin{align}\label{eqn:uplink signal}
y_{m,n}^{{\rm UL}}=\sum_{u=1}^M \sum_{l=0}^{L+\epsilon_{\max}-1} \tilde{h}^{\rm UU}_{u,m,l} \bar{\gamma}_{u,n-l} + \hat{z}_{m,n}, \forall m,n,
\end{align}where $\epsilon_{\max} = \max_{u,m}|\epsilon_{u,m}|$ denotes the maximum absolute STO among the UEs and $\tilde{h}^{\rm UU}_{u,m,l}$ is defined as 
\begin{equation}\label{tildehuu}
\tilde{h}_{u,m,l}^{\rm UU}=\begin{cases}
	&h_{u,m,l-\epsilon_{u,m}}^{\rm UU}, ~  \text{if}~l \in [l_{u,m}+\epsilon_{u,m},L+\epsilon_{u,m}-1],\\
	&0, \quad \quad \quad \quad ~ ~ \text{otherwise}.
\end{cases}
\end{equation}  

According to (\ref{eqn:uplink signal}), the frequency domain signals of UE $m$ received at its own sub-carriers $\mathcal{N}_u^{(m)}$ can be given as
\begin{align}
\bar{\boldsymbol{y}}_m^{{\rm UL}}=\sqrt{p_m} \diag(\boldsymbol{r}_m)\boldsymbol{E}_m\tilde{\boldsymbol{h}}^{\rm UU}_{m} + \bar{\boldsymbol{z}}_m^{\rm UL},
\end{align} where $\tilde{\boldsymbol{h}}_{m}^{\rm UU}=[\tilde{h}_{u,m,0}^{\rm UU},\dots,\tilde{h}_{u,m,L+\epsilon_{\max}-1}^{\rm UU}]^T$ denotes the virtual channel vector between UE $u$ and UE $m$ with length $L+\epsilon_{\max}$, $\boldsymbol{E}_m \in \mathbb{C}^{|\mathcal{N}_u^{(m)}|\times (L+\epsilon_{\max})}$ with the $(n,l)$-th element being $E_{n,l}=e^{\frac{-j2\pi(n-1)(l-1)}{N}}$, and $\bar{\boldsymbol{z}}_m^{\rm UL} \sim \mathcal{CN}(0,\hat{\sigma}_m^2 \boldsymbol{I}_{|\mathcal{N}_u^{(m)}|})$ denotes the noise at UE $m$ over its assigned sub-carriers.

Based on the same method proposed for the downlink case as shown in (\ref{pu})-(\ref{Dbtu}) of Section \ref{sec:Active UE based Sensing Mode}, each UE $m$ can first estimate the time domain channels $\tilde{\boldsymbol{h}}_m^{{\rm UU}}$ based on $\bar{\boldsymbol{y}}_m^{{\rm UL}}$ and then estimate range information based on the non-zero elements in the channels $\tilde{\boldsymbol{h}}_m^{{\rm UU}}$. Specifically, let $\mathcal{D}_m^{{\rm UT}}$ denote the set of distance values obtained by UE $m$ from the uplink echo signals (\ref{u_signal}), similar to $\mathcal{D}_m^{{\rm BTU}}$ given in (\ref{Dbtu}) obtained from the downlink echo signals (\ref{bs2ue}).   

In Phase II, our goal is to modify Algorithm \ref{alg2} via utilizing the additional information about $\mathcal{D}_m^{{\rm BTU}}$'s such that the performance can be improved while the complexity can be reduced. Note that $\mathcal{D}_m^{{\rm UT}}$ consists of the distance values from UE $m$ to the targets. However, we still have the data association issue, i.e., we do not know which element in $\mathcal{D}_m^{{\rm UT}}$ is an estimation of $d_{m,1}^{\rm UT}$, which element is an estimation of $d_{m,2}^{\rm UT}$, and so on. Similar to the passive UE based sensing mode shown in Section \ref{sec:Phase II: Target Localization via Joint Data Association and Ineffective Anchors Removing}, define $e_{m,k}$ as a data association integer such that the estimated distance between UE $m$ and target $k$, denoted by $\bar{d}^{\rm UT}_{m,k}$, is the $e_{m,k}$-th largest element in $\mathcal{D}^{\rm UT}_m$, $\forall m, k$, i.e., $\bar{d}^{\rm UT}_{m,k} = \mathcal{D}^{\rm UT}_m(e_{m,k})$. 
Define
\begin{align}\label{diffe}
& \gamma_{m,k}(g_{m,k},e_{m,k}) \nonumber \\ =&\left |\mathcal{D}^{\rm BTU}_m(g_{m,k})-\mathcal{D}^{\rm BT}(k)-\mathcal{D}^{\rm UT}_m(e_{m,k}) \right|, ~ \forall m,k. 
\end{align}
Given any data association solution $g_{m,k}$ and $e_{m,k}$, $\gamma_{m,k}(g_{m,k},e_{m,k})$ denotes the estimation residue based on \eqref{sumdistance}.

Compared to the passive UE based sensing mode discussed in Section \ref{sec:Localization Method under Passive UE based Sensing Mode}, the estimation residue shown in (\ref{diffe}) is the new information arising from the sets $\mathcal{D}^{\rm UT}_m$'s obtained under the active UE based sensing mode. The roles of the estimation residue given in (\ref{diffe}) are two-fold. First, (\ref{diffe}) is useful for removing the ineffective UEs. Specifically, we rely on the LOS links among the BS and the UEs to correct the synchronization errors, e.g., (\ref{eqn:estiamted LOS}), (\ref{lm}), and (\ref{taum}) can be used to synchronize the BS with the UEs. However, such a synchronization method relies on whether (\ref{eqn:estiamted LOS}) is a good approximation of (\ref{eqn:LOS delay}). For example, if UE $m$'s estimated position is perfect, i.e., $\hat{\boldsymbol{u}}_m=\boldsymbol{u}_m$, then $\tilde{l}_m=l_m$. Thereby, the synchronization between the BS and UE $m$ is quite perfect, such that the negligible STO will not affect the estimation of $d_{m,k}^{\rm BTU}$'s, $k=1,\ldots,K$. In this case, given the correct data association solution $g_{m,k}$'s and $e_{m,k}$'s, $\gamma_{m,k}(g_{m,k},e_{m,k})$'s can be small, $\forall k$. However, if UE $m$'s estimated position is quite erroneous, then $\tilde{l}_m$ is a bad approximation of $l_m$. In this case, the STO between the BS and UE $m$ is not accurately estimated, and it will significantly affect the delay and range estimation, i.e., $\gamma_{m,k}(g_{m,k},e_{m,k})$'s can be large, $\forall k$, even given the correct data association solution $g_{m,k}$'s and $e_{m,k}$'s. To summarize, the estimation residues $\gamma_{m,1}(g_{m,1},e_{m,1}),\ldots,\gamma_{m,K}(g_{m,K},e_{m,K})$ can reflect whether UE $m$ is an ineffective UE or not, $\forall m$, given the right data association solution. Second, the estimation residues $\gamma_{m,1}(g_{m,1},e_{m,1}),\ldots,\gamma_{m,K}(g_{m,K},e_{m,K})$ can also reflect whether the data association solution $g_{m,1},\ldots,g_{m,K}$ and $e_{m,1},\ldots,e_{m,K}$ is correct or not, because if $\mathcal{D}^{\rm BTU}_m(g_{m,k})$ and $\mathcal{D}^{\rm UT}_m(e_{m,k})$ do not belong to target $k$, $\gamma_{m,k}(g_{m,k},e_{m,k})$ can be large.

In the following, we first show how to find part of the ineffective UEs and then introduce how to narrow the feasible regions for the data association variables $g_{m,k}$'s based on the above observations, respectively.
Specifically, let $g_{m,k}^\ast$ and $e_{m,k}^\ast$ denote the true solutions to $g_{m,k}$ and $e_{m,k}$, respectively. Based on the above discussion, if UE $m$ is effective, then the corresponding residue defined in \eqref{diffe}, i.e., $\gamma_{m,k}(g_{m,k}^\ast, e_{m,k}^\ast)$, should be small and satisfy
\begin{equation}\label{gammath}
    \gamma_{m,k}(g_{m,k}^\ast, e_{m,k}^\ast)  \leq \gamma_{\rm th}, \forall k,
\end{equation}
where $\gamma_{\rm th}$ is a predetermined threshold. 
We further define 
\begin{align}
   \mathcal{E}_{m,k} = \left\{(g_{m,k},e_{m,k})|g_{m,k} \in \mathcal{K}, e_{m,k} \in \mathcal{K} \right\}
\end{align}
as the set of all the possible solutions to $(g_{m,k},e_{m,k})$, $\forall m, k$. Then, given each element in $\mathcal{E}_{m,k}$, we can obtain a residue as defined in \eqref{diffe}. For conveniences, define the minimum residue as $\gamma_{m,k}^{\min}$ as 
\begin{equation}
	\gamma_{m,k}^{\min} = \underset{(g_{m,k},e_{m,k}) \in \mathcal{E}_{m,k}}{\min} \gamma_{m,k}(g_{m,k},e_{m,k}).
\end{equation}
 Since $(g_{m,k}^\ast,e_{m,k}^\ast) \in \mathcal{E}_{m,k}$, we have
\begin{equation}\label{conforM}
    \gamma_{m,k}^{\min} \leq \gamma_{m,k}(g_{m,k}^\ast, e_{m,k}^\ast)  \leq \gamma_{\rm th}, \forall k.
\end{equation}
Then, we can obtain a necessary condition for UE $m$ to be effective: if UE $m$ is effective, then $\gamma_{m,k}^{\min} \leq \gamma_{\rm th}$ must hold given each target $k$, i.e., 
\begin{equation}\label{uecons}
    \gamma_{m,k}^{\min} \leq \gamma_{\rm th}, \forall k.
\end{equation}
In other words, if there exists a target $k$ such that $\gamma_{m,k}^{\min} \leq \gamma_{\rm th}$ does not hold, then UE $m$ is an ineffective UE, the set of which is defined as
\begin{equation}\label{barM}
	\bar{\mathcal{M}} = \{m|\exists k, \gamma_{m,k}^{\min} > \gamma_{\rm th} \}.
\end{equation}
Moreover, based on \eqref{gammath}, for each effective UE $m$, the feasible solution to $(g_{m,k},e_{m,k})$ should satisfy
\begin{equation}\label{th4ge}
    \gamma_{m,k}(g_{m,k}, e_{m,k})  \leq \gamma_{\rm th}.
\end{equation}
Then, the feasible region for $(g_{m,k},e_{m,k})$ can be narrowed to
\begin{equation}\label{bare}
   \!\! \bar{\mathcal{E}}_{m,k}\! =\! \left\{(g_{m,k},e_{m,k})|g_{m,k} \in \mathcal{K}, e_{m,k} \in \mathcal{K}, \eqref{th4ge}~\text{holds} \right\}.
\end{equation}
We further define $\bar{\mathcal{G}}_{m,k}$ as the set containing all the solutions to $g_{m,k}$ in $\bar{\mathcal{E}}_{m,k}$. Then, $\bar{\mathcal{G}}_{m,k}$ is the narrowed feasible region to $g_{m,k}$, i.e., $g_{m,k} \in \bar{\mathcal{G}}_{m,k}, \forall k$.

Because the estimation residues given in (\ref{diffe}) can help on identifying the ineffective UEs and wrong data association solutions, we can modify Algorithm \ref{alg2} as follows to make it better under the active UE based sensing mode. 
First, at the beginning of Algorithm~\ref{alg2}, we can first remove the ineffective UEs in $\bar{\mathcal{M}}$ defined in \eqref{barM} from $\mathcal{M}$ to obtain a new UE set $\hat{\mathcal{M}} \triangleq \mathcal{M} \setminus \hat{\mathcal{M}}$. Then, in the following steps in Algorithm~\ref{alg2}, we only need to utilize the position and range information of the UEs in $\hat{\mathcal{M}}$ to localize the targets.   Second, with the knowledge about the new feasible region about $g_{m,k}$, i.e., $\bar{\mathcal{G}}_{m,k}$, the feasible regions for $\tilde{\mathcal{G}}_k$ and $\tilde{\mathcal{G}}_{k+1}$ used in step 1 and step 4 of Algorithm~\ref{alg2}, i.e., ${\mathcal{H}}_k$ and ${\mathcal{H}}_{k+1}$, can be respectively replaced by
\begin{equation}\label{newdak}
    \tilde{\mathcal{H}}_k = \left\{\tilde{\mathcal{G}}_k|g_{m,k} \in \bar{\mathcal{G}}_{m,k}, \forall m \in \mathcal{M} \right\},
\end{equation}
\vspace{-4mm}
\begin{equation}\label{newdak1}
\begin{aligned}
    &\tilde{\mathcal{H}}_{k+1} \\
    &= \left\{{\mathcal{G}}_{k+1}|g_{m,k+1}\!\in\! \bar{\mathcal{G}}_{m,k+1}, g_{m,k+1}\! \neq\! g_{m,k}^\ast, \forall m \! \in\! \tilde{\mathcal{M}}(\tilde{\mathcal{G}}_k^\ast) \right\}.
\end{aligned}
\end{equation}
Since some ineffective UEs have been removed and there are significantly fewer elements in $\tilde{\mathcal{H}}_k$ and $\tilde{\mathcal{H}}_{k+1}$ compared to ${\mathcal{H}}_k$ and ${\mathcal{H}}_{k+1}$, the performance of Algorithm~\ref{alg2} can be enhanced in terms of both localization accuracy and computational complexity after the above operations. 
  \vspace{-2mm}
\section{Numerical Results}\label{sec:Numerical Results}
In this section, we provide numerical results to verify the effectiveness of the proposed STO estimation method and the proposed two-phase localization protocol. In our numerical examples, the BS, the UEs, and the targets are uniformly distributed in a 100 m $\times$ 100 m square. The position uncertainty of UE $m$, i.e., $\Delta \boldsymbol{u}_m$, is modeled as a zero-mean Gaussian random vector with covariance matrix $\boldsymbol{\Phi}_m, \forall m$ \cite{ghasem2014}.  Specifically, if UE $m$ is effective, we set $\boldsymbol{\Phi}_m = -20 \boldsymbol{I}_2 ~\text{dBm}^2$. Otherwise, we set $\boldsymbol{\Phi}_m = 20 \boldsymbol{I}_2 ~\text{dBm}^2$. 
Furthermore, in problem (P1-$k$), we set $v = 10$. Unless otherwise stated, the transmit powers of the BS and each UE are 20 Watt (W) and 2 W, respectively. 
The bandwidth of the downlink signals is 400 MHz, while that of the uplink signals, denoted by $B_u$, is either 20 MHz or 100 MHz \cite{38101}. Then, we introduce how to generate the  multi-path channels, i.e., $\boldsymbol{h}^{\rm BTB}$, $\boldsymbol{h}_m^{\rm BU}$'s, and $\boldsymbol{h}_{u,m}^{\rm UU}$'s, in numerical examples. For simplicity, we mainly show how to generate $\boldsymbol{h}_m^{\rm BU}$'s, while $\boldsymbol{h}^{\rm BTB}$ and $\boldsymbol{h}_{u,m}^{\rm UU}$'s can be generated in the similar way. 
Specifically, there are $K+1$ non-zero elements in $\boldsymbol{h}_m^{\rm BU}$ comprising the LOS channel from the BS to UE $m$ and the $K$ channels from the BS to the targets to UE $m$, $\forall m$. 
For a LOS channel of a path with range $d$ meters, the path loss model is $\beta_0(\frac{d}{d_0})^{-\alpha}$, where $d_0=1$ m denotes the reference distance, $\beta_0=-20$ dB denotes the path loss at the reference distance, and $\alpha=2$ denotes the path loss factor\cite{Wang2020RIS}. Then, the LOS BS-UE channels can be generated. Moreover, the radar cross section (RCS) of the targets is set as -10 $\text{dBm}^2$. Then, each BS-target-UE channel is modeled by the product among the BS-target path loss, the target RCS, and the target-UE path loss.

\vspace{-5mm}
\subsection{Performance of the STO Estimation Method in Phase I}\label{subsec:sto estimation}
\begin{figure}[t]
	\centering
	\includegraphics[height=5.8cm]{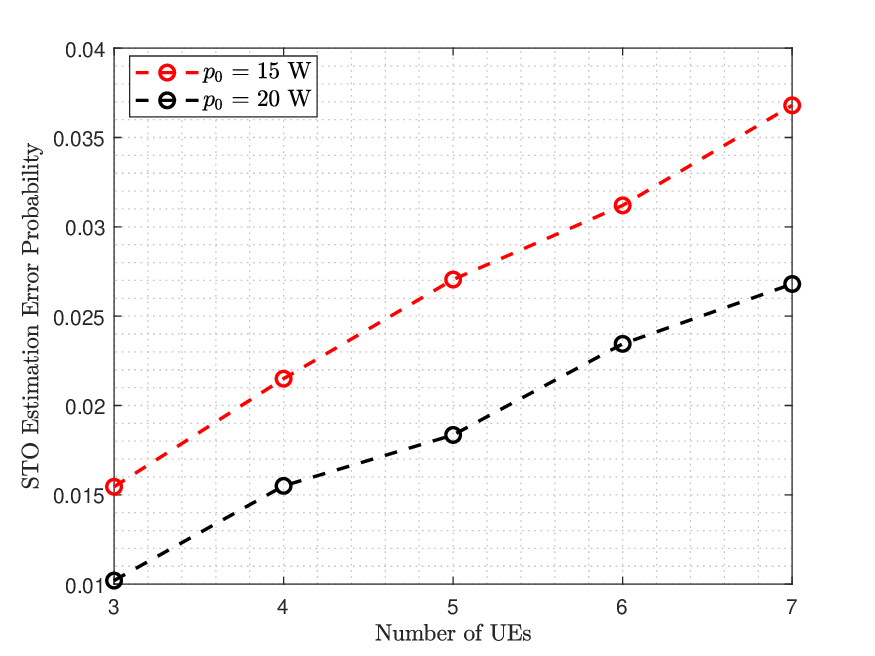}
	\caption{STO estimation error probability versus the number of UEs.} \label{sto}
	\vspace{-5mm}
\end{figure}
In this subsection, we evaluate the performance of the proposed STO estimation method in Phase I. Specifically, we set $N_d = 3300$ and $\Delta f_d = 120$ kHz such that the downlink bandwidth is 400 MHz \cite{38104}. In this case, the length of the CP is $0.59 \mu$s \cite{Zaidi2017}. Moreover, the maximum absolute STO among the BS and the UEs is set to 10 OFDM sample periods, i.e., $\tau_{\max} = 10$, and the STO between the BS and UE $m$, i.e., $\tau_m$, is randomly generated in the interval $[-\tau_{\max},\tau_{\max}]$. To guarantee $Q_d>L+\tau_{\max}$ such that all the ISI is received within the CP, the maximum number of resolvable paths is $L = 200$ \cite{Shi2024}. 
Note that we only consider effective UEs in this subsection since the ineffective UEs cannot accurately estimate their STOs as illustrated before. We generate $10^5$ Monte Carlo experiments. In each experiment, we randomly generate the locations of the BS, the UEs, and the targets, and implement STO estimation. In each experiment, if there is at least one UE whose STO is not accurately estimated, then this experiment is regarded as unsuccessful. 
Define the STO estimation error probability as the ratio between the number of unsuccessful experiments and the total number of experiments. Fig.~\ref{sto} shows the STO estimation error probability versus the number of UEs in a network consisting of 1 BS and 4 targets when the BS transmits with 15 W and 20 W, respectively. It is observed that the STO estimation error probability of the proposed method is quite low. Moreover, it is observed that the STO estimation error probability increases with the number of UEs. The reason for this is as follows. The STO estimation error occurs when at least one UE's STO is incorrectly estimated. Therefore, with a larger number of UEs, the probability of encountering at least one UE with an inaccurately estimated STO increases, resulting in a higher STO estimation error probability.
\subsection{Performance of Passive UE and Active UE based Sensing Modes with Single Target}
In this subsection, we evaluate the performance of the proposed two-phase localization protocol under the passive UE based and the active UE based sensing modes with single target. For comparison, we adopt the following benchmark scheme:
\begin{itemize}
	\item \textit{Benchmark Scheme I-S:} Under this scheme, range estimation is the same as our method in Section III. In Phase II, we do not perform UE selection but directly use all the UEs as anchors to localize the targets based on the multilateration method \cite{Torrieri1984}.   
\end{itemize}

In this numerical example, we generate $10^5$ Monte Carlo experiments. In each experiment, we randomly generate the positions of the BS, the UEs, and the target in the considered area, and localize the target using the proposed scheme and the benchmark scheme. Here, an error event for localizing a target is defined as the case that the estimated location of the target does not lie within a radius of 1 m from the true target location. Let $N_m$ denote the total number of error events in these $10^5$ experiments. Then, the localization error probability is defined as $\frac{N_m}{K\times 10^5}$.

\begin{figure}[t]
	\centering
	\includegraphics[height=5.8cm]{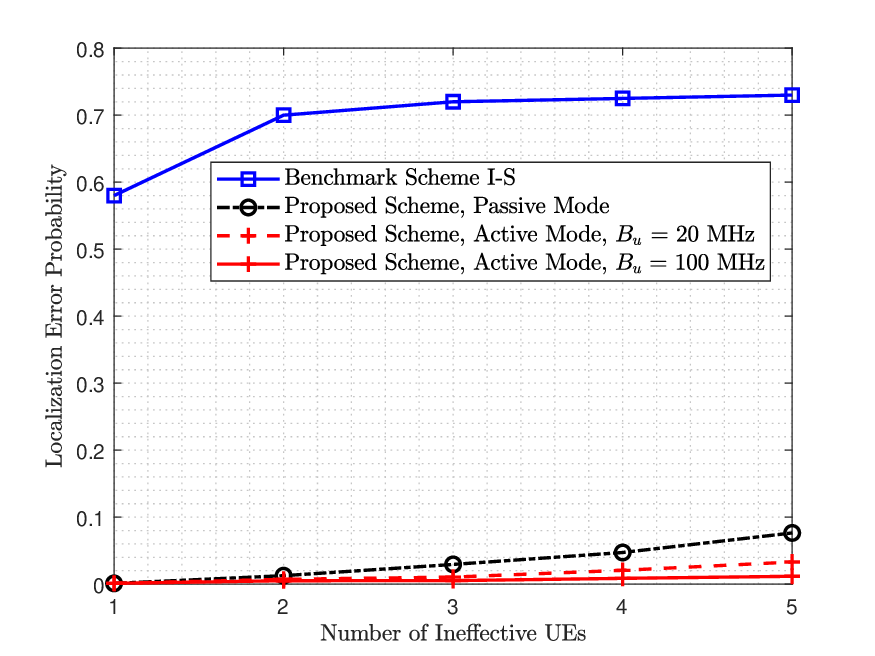}
	\caption{Performance comparison of the benchmark scheme and the proposed scheme under the passive UE based and the active UE based sensing modes with single target.}\label{fig:singlerr}
	\vspace{-5mm}
\end{figure}

Fig.~\ref{fig:singlerr} shows the performance of our proposed scheme and Benchmark Scheme I-S with 4 effective UEs and the number of the ineffective UEs ranging from 1 to 5. Compared to Benchmark Scheme I-S, we can see that the proposed scheme under both passive UE based and active UE based sensing modes can achieve high-accuracy localization of the target via removing the ineffective UEs from the anchor set.

\subsection{Performance of Passive UE based Sensing Mode with Multiple Targets}
In this subsection, we evaluate the performance of the proposed two-phase localization protocol under the passive UE based sensing mode with multiple targets. For comparison, we consider the following benchmark schemes:
\begin{itemize}
	\item \textit{Benchmark Scheme I:} Under this scheme, we do not perform STO compensation in Phase I. In this case, the estimated ranges at the UEs based on the downlink signals are corrupted by the STOs. In phase II, we localize the targets using the proposed method. 
	\item \textit{Benchmark Scheme II:} Under this scheme, range estimation is the same as the proposed method in Section~\ref{sec:Localization Method under Passive UE based Sensing Mode} while in Phase II, we do not perform UE selection but directly use all the UEs as anchors to localize the targets based on the method in \cite{shi2022}. 
	\item \textit{Benchmark Scheme III:} Under this scheme, range estimation is the same as our method in Section~\ref{sec:Localization Method under Passive UE based Sensing Mode}. In Phase II, we first localize target 1 via solving problem (P5-1-M) and obtain a set of
	UEs, i.e., $\tilde{\mathcal{M}}(\tilde{\mathcal{G}}_1^\ast)$. 
	Then, we do not check the effectiveness of
	these UEs via localizing target 2 but directly use them as anchors to localize the other targets using the method in \cite{shi2022}.
	\item \textit{Benchmark Scheme IV:} Under this scheme, range estimation is the same as our method in Section~\ref{sec:Localization Method under Passive UE based Sensing Mode}. In Phase II, we assume that the effective UEs and data association are perfectly known and localize the targets based on the multilateration method \cite{Torrieri1984}. This scheme can serve as an error probability lower bound. 	
\end{itemize}
\begin{figure}
	\centering
	\includegraphics[height=5.8cm]{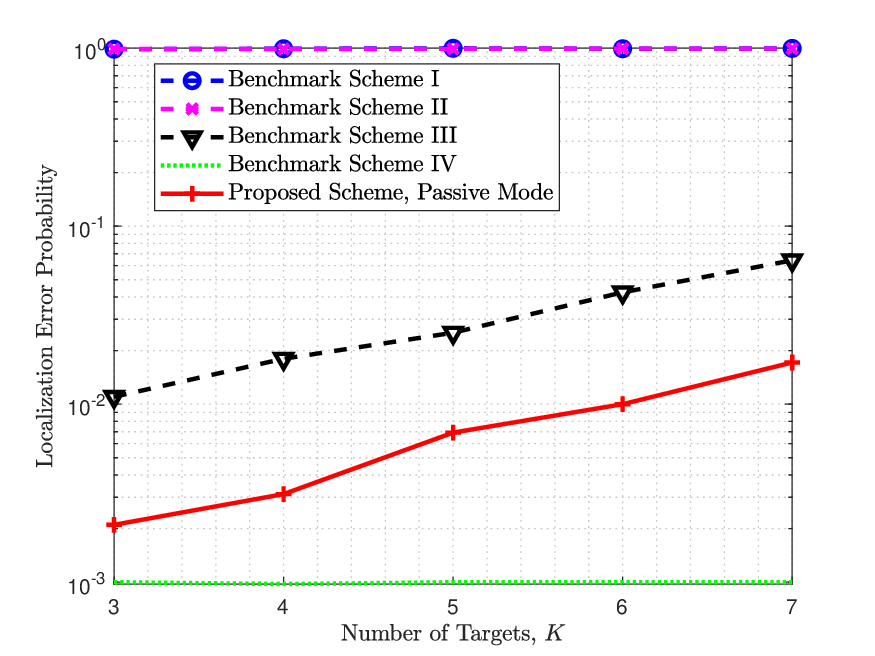}
	\caption{Performance comparison of the benchmark schemes and the proposed scheme under the passive UE based sensing mode with multiple targets.} \label{fig:err1}
	\vspace{-5mm}
\end{figure}

Fig.~\ref{fig:err1} shows the performance comparison between the proposed scheme and the benchmark schemes in terms of localization error probability in a scenario with 3 effective UEs and 2 ineffective UEs.
First, based on the performance of Benchmark Scheme I and Benchmark Scheme II, we can see that the existence of STOs and ineffective UEs can lead to unacceptable localization error probability. 
Then, it is observed that the proposed scheme can achieve high-accuracy localization of the targets by compensating the STOs in Phase I and jointly optimizing UE selection and data association in Phase II. 
It is also observed that the proposed scheme outperforms Benchmark Scheme III. The reason is that we can find the effective UEs with a higher success rate by rechecking the effectiveness of the obtained UEs via localizing different targets as in the proposed scheme than by just localizing target 1 as in Benchmark Scheme III. 
Moreover, there is a small performance gap between the proposed scheme and Benchmark Scheme IV, where the effective UEs and data association are perfectly known.  
 
\subsection{Comparison between the Passive UE and Active UE based
	Sensing Modes with Multiple Targets}

\begin{figure}[t]
 	\centering
 	\subfigure[Localization performance]{\includegraphics[height=5.8cm]{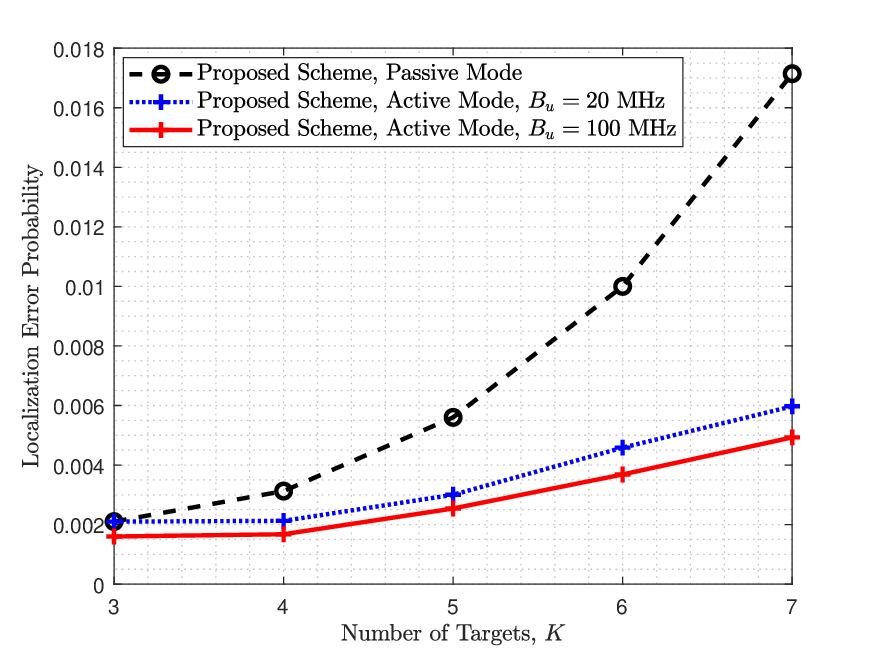}\label{fig:err2}}\\[-1mm]
 	\subfigure[CPU Time]{\includegraphics[height=5.8cm]{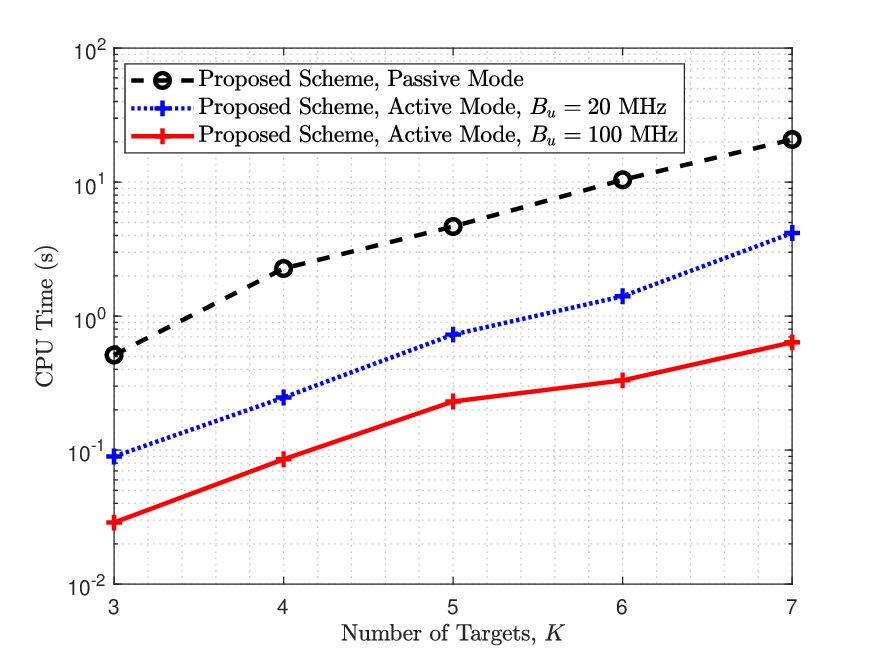}\label{fig:time2}}
 	\caption{Performance comparison of the proposed scheme under the passive UE based and the active UE based sensing modes with multiple targets.} \label{exp2}
 	\vspace{-5mm}
 \end{figure}

Next, we aim to compare the passive UE and the active UE based sensing modes in terms of performance and complexity in the multi-target case. Specifically, Fig.~\ref{fig:err2} shows the performance comparison between the proposed scheme under the passive UE based and the active UE based sensing modes.
It is observed that the proposed scheme under the active mode achieves a lower localization error probability compared to the passive mode. The reasons for this are as follows. First, under the active UE based sensing mode, we can remove some ineffective UEs based on \eqref{barM} before we perform localization. With fewer ineffective UEs, we are able to find the effective UEs with a higher probability of success, resulting in a lower localization error probability. Second, based on the ranges obtained via uplink signals, we can eliminate many infeasible data association solutions for each target based on \eqref{bare}. This can also reduce the localization error probability.       

Fig.~\ref{fig:time2} shows the average CPU running time to implement the proposed localization method under different sensing modes. We can see that the time cost under the active mode is significantly lower than that under the passive mode since we can remove some ineffective UEs and reduce the size of the set consisting of the feasible data association solutions thanks to the additional range information obtained via uplink signals. 

Moreover, by comparing the performance of the proposed scheme under the active mode with different uplink bandwidths, we observe that a higher uplink channel bandwidth brings higher localization accuracy and lower localization complexity.
This is because if the uplink bandwidth is larger, the range information obtained by the UEs via the uplink signals is more accurate. Then, with this more accurate side information, more ineffective UEs and wrong data association solutions can be removed based on \eqref{barM} and \eqref{bare}, leading to higher localization accuracy and lower computational time to find all the ineffective UEs and correct data association solution.  

\vspace{-2mm}
\section{Conclusions}\label{sec:Conclusions}
In this paper, we investigated the UE assisted sensing framework for 6G ISAC. Under this scheme, the UEs estimate the range information of the targets and share these information with the BS, which localizes the targets via fusing the global sensing information. In practice, the UEs are imperfectly synchronized and with erroneous location information from the GPS. To enable UE assisted sensing, we proposed an efficient algorithm that can jointly estimate the STOs between the BS and the UEs and remove the UEs with quite erroneous position information from the anchor set. Future work may explore the possibility of UE assisted sensing in other applications such as detection, tracking, etc. 

\bibliographystyle{IEEEtran}
\bibliography{ref}

\end{document}